\documentclass[journal]{IEEEtran}
\usepackage{amssymb}
\usepackage{amsbsy}
\usepackage{amsmath}
\usepackage{amsthm} 
\usepackage{setspace}
\usepackage{epsfig}
\usepackage{cite}
\usepackage{graphics}
\usepackage{epstopdf}
\usepackage{mathrsfs}
\usepackage[above]{placeins}
\usepackage{algorithm}
\usepackage{algpseudocode}
\usepackage{multirow}
\usepackage{algorithmicx}
\usepackage{tikz}

\begin{document}

%

\title{\huge{Markov chain Monte Carlo Methods For Lattice Gaussian Sampling: Lattice Reduction and Decoding Optimization}}

\author{Zheng~Wang,~\IEEEmembership{Member, IEEE,}
        Yang~Huang,~\IEEEmembership{Member, IEEE,}
        and Shanxiang~Lyu

\thanks{

This work was supported in part by the National Natural Science Foundation of China under Grant 61801216, in part by the Natural Science Foundation of Jiangsu Province
under Grant SBK2018042902.

Z. Wang and Y. Huang are with College of Electronic and Information Engineering, Nanjing University of Aeronautics and Astronautics (NUAA), Nanjing, China; S. Lyu is with the Department of Electrical and Electronic Engineering, Imperial
College London, London, SW7 2AZ, United Kingdom (e-mail: z.wang@ieee.org, yang.huang.ceie@nuaa.edu.cn, s.lyu14@imperial.ac.uk).


}}

\maketitle


\begin{abstract}
Sampling from the lattice Gaussian distribution has emerged as an important problem in coding, decoding and cryptography.
In this paper, lattice reduction technique is adopted to Gibbs sampler for lattice Gaussian sampling.
Firstly, with respect to lattice Gaussian distribution, the convergence rate of systematic scan Gibbs sampling is derived and we show it is characterized by the Hirschfeld-Gebelein-R\'enyi (HGR) maximal correlation among the multivariate of being sampled. Therefore, lattice reduction is applied to formulate an equivalent lattice Gaussian distribution but with less correlated multivariate, which leads to a better Markov mixing due to the enhanced convergence rate. Then, we extend the proposed lattice-reduction-aided Gibbs sampling to lattice decoding, where the choice of the standard deviation for the sampling is fully investigated.
A customized solution that suits for each specific decoding case by Euclidean distance is given, thus resulting in a better trade-off between Markov mixing and sampler decoding. Moreover, based on it, a startup mechanism is also proposed for Gibbs sampler decoding, where decoding complexity can be reduced without performance loss.
Simulation results based on large-scale MIMO detection are presented to confirm the performance gain and complexity reduction.
\end{abstract}

\IEEEpeerreviewmaketitle

\vspace{0.5em}


\IEEEpeerreviewmaketitle

\section{Introduction}
\IEEEPARstart{N}{o}wadays, lattice Gaussian sampling has drawn a lot of attention in various research fields. In mathematics, Banaszczyk was the first to apply it to prove the transference theorems for lattices \cite{Banaszczyk}. In coding,
lattice Gaussian distribution was employed to obtain the full shaping gain for lattice coding \cite{Forney_89,Kschischang_Pasupathy,LiuLing2}, and to achieve the capacity of the Gaussian channel \cite{LB_13}. It was also used to achieve information-theoretic security in the Gaussian wiretap channel \cite{LiuLing1,LLBS_12,7360779} and in the bidirectional relay channel \cite{7058433}, respectively. In cryptography, the lattice Gaussian distribution has become a central tool in the construction of many primitives \cite{MicciancioGaussian,Regevlearning,GentryDissertation}. Specifically, lattice Gaussian sampling lies at the core of signature schemes in the Gentry, Peikert and Vaikuntanathan (GPV) paradigm \cite{Trapdoor}.
In decoding, lattice Gaussian sampling with a suitable variance allows to solve the closest vector problem (CVP) and the shortest vector problem (SVP) \cite{RegevSolvingtheShortestVectorProblem,RegevSolvingtheClosestVectorProblem}.
From the viewpoint of lattice decoding, the optimal maximal likelihood (ML) detection in multiple-input multiple-output (MIMO) systems corresponds to solving the CVP \cite{DamenDetectionSearch}.

However, in sharp contrast to the continuous Gaussian density, it is by no means trivial even to sample from a low-dimensional discrete Gaussian distribution. Efficient sampling schemes do exist but they only work for a few special lattices \cite{LB_13,CB16}.
As the default sampling algorithm for general lattices, Klein's algorithm \cite{Klein} only works when the standard deviation $\sigma = \sqrt{\omega(\text{log}\ n)}\cdot\text{max}_{1\leq i \leq n}\|\mathbf{\widehat{b}}_i\|$ \cite{Trapdoor}, where $\omega(\text{log}\ n)$ is a superlogarithmic function, $n$ denotes the lattice dimension and $\mathbf{\widehat{b}}_i$'s are the Gram-Schmidt vectors of the lattice basis $\mathbf{B}$.
Unfortunately, such a requirement of $\sigma$ tends to be excessively large, rendering Klein's algorithm inapplicable to many scenarios of interest.

%
%

In order to sample from a target lattice Gaussian distribution with arbitrary $\sigma>0$, Markov chain Monte Carlo (MCMC) methods were introduced in \cite{ZhengWangTIT15,ZhengWangTIT17}. In principle, it randomly generates the next Markov state conditioned on the previous one; after the burn-in time, which is normally measured by the \emph{mixing time}, the Markov chain will step into a stationary distribution, when samples from the target distribution can be obtained \cite{mixingtimemarkovchain}.
As a basic MCMC method, the Gibbs sampling, which employs univariate conditional sampling to build the Markov chain, has been introduced to lattice Gaussian sampling by showing its ergodicity\cite{ZhengWangMCMCLatticeGaussian}. In \cite{ZhengWangSysmetricGibbs}, the symmetric Metropolis-within-Gibbs (SMWG) algorithm was proposed for lattice Gaussian sampling to achieve the exponential convergence. Moreover, the Markov chain induced by random scan Gibbs sampling for lattice Gaussian distribution was shown to be geometric ergodicity \cite{ITW2017}, which means it converges exponentially fast to the stationary distribution.

On the other hand, with the increment of antenna numbers, the large-scale MIMO system has become a promising extension of MIMO, which boosts the network capacity on a much greater scale. The dramatically increased system size places a pressing challenge on the signal detection while a lot of research attentions have been attracted by it \cite{TabuSrinidhi,DaiL1,GaoX1,MSMIMO1,ZZY1}.
Thanks to the \emph{convergence theorem} of MCMC, Gibbs sampling with a finite state space naturally experiences the geometric ergodicity, so that it has already been adapted to MIMO detection to solve the CVP \cite{HassibiMCMCnew,McmcDatta,XiaodongWangMultilevel,MCMCHaidongZhu,ChoiMCMC1}.
Meanwhile, Gibbs sampling has also been introduced into soft-output decoding in MIMO systems, where the extrinsic information calculated by a priori probability (APP) detector is used to produce soft outputs \cite{MCMCBehrouz,BaiLin1}. In \cite{ChenMCMC}, an investigation of Gibbs-based MCMC receivers in different communication channels is given as well.

However, given those works, the choice of the standard deviation $\sigma$ (also referred to as ``temperature'') for Gibbs sampling decoding has not been fully investigated. A common choice comes from statistics by letting $\sigma^2$ be the variance of noises, which severely suffers from the \emph{stalling problem} in high signal-to-noise ratio (SNR) regime. Although Hassibi \textit{et al.} suggested $\sigma$ should instead be scaling at least as $\Omega(\sqrt{\text{SNR}})$, it fails to exploit the decoding potential for each specific case \cite{HassibiMCMCnew}.


Meanwhile, another very important point was ignored for years. Specifically, as an advanced decoder, Gibbs sampler decoding, however, is not necessary for all the decoding cases, where the optimal solution may be directly obtained by suboptimal decoding schemes especially in high SNRs. This indicates substantial computational complexity can be saved without any performance loss. In \cite{McmcDatta,ChoiMCMC2}, two stopping criterions were given for mixed-Gibbs sampler decoding schemes, but they only work for the proposed multiple restart strategies by terminating those trapped Markov chains.


In this paper, we advance the state of the art of the MCMC-based lattice Gaussian sampling in several fronts. First of all, in order to enhance the convergence performance of Gibbs sampling, the lattice-reduction-aided Gibbs sampling algorithm is proposed for lattice Gaussian sampling. In particular, a comprehensive analysis regarding to the convergence rate of the Markov chain induced by systematic scan Gibbs sampling is presented, and we show the convergence is essentially dominated by the Hirschfeld-Gebelein-R\'enyi (HGR) maximal correlation between the multiple random variables.
Hence, by lattice reduction, an equivalent lattice Gaussian distribution can be established with significantly reduced HGR maximal correlation, thus leading to a boosting convergence performance.
Note that the block strategy, which improves the convergence by performing the sampling over multivariate within one Markov move \cite{ZhengWangMCMCLatticeGaussian}, also works in the proposed lattice-reduction-aided Gibbs sampling.

We then extend the lattice-reduction-aided Gibbs sampling algorithm to lattice decoding. The investigation starts from optimizing the sampling probability of the target decoding point, which leads to a better trade-off between Markov mixing and sampling decoding. Specifically, we show that the choice of the standard deviation $\sigma$ heavily depends on the distance from the query point to the lattice. This not only effectively avoids the stalling problem, but also provides a preferable choice of $\sigma$ for each specific decoding case. Furthermore, for a better approximation of $\sigma$, the initial starting point of the Markov chain is strongly desired to be well chosen while this is actually in accordance with geometric ergodicity as the initial starting point also has an indispensable impact on the convergence behaviour.
Then, based on the initial starting point, we adopt the correct decoding radius from bounded distance decoding (BDD) to build a startup mechanism, which decides whether to invoke Gibbs sampler or not.
Meanwhile, the demand of the high quality initial starting point can also be guaranteed through the usage of lattice reduction. In a word, our proposed Gibbs sampler decoding advances with better decoding performance and less complexity cost.


It should be noticed that compared to the lattice Gaussian distribution, the discrete Gaussian distribution designed for MIMO detection entails a finite state space (i.e., $\mathbf{x}\in\mathcal{X}^n$ based on the QAM constellation). After the nonlinear transformation $\mathbf{z}=\mathbf{U}^{-1}\mathbf{x}$ of lattice reduction ($\mathbf{U}\in \mathbb{Z}^{n\times n}$ is a unimodular matrix with $\text{det}(\mathbf{U})=\pm1$), the state space of $\mathbf{z}$ turns out to be computationally expensive to get.
For non-Gibbs sampling based detectors \cite{WubbenMMSE}, suboptimal remedies can be carried out to restrict $\widehat{\mathbf{x}}=\mathbf{Uz}$ to the original set $\mathcal{X}^n$ in the end. However, for Gibbs sampler decoding, the Markov chain along with an unbounded or approximate state space of $\mathbf{z}$ tends to be unreasonably wild, which most likely results in an invalid Markov mixing. Such a problem does not exist in lattice decoding paradigm since $\mathbf{x}$ and $\mathbf{z}$ share the same state space $\mathbb{Z}^n$. To this end, lattice reduction is not recommended to be directly applied in the Markov mixing of Gibbs sampling for MIMO detection.
Nevertheless, the aforementioned analysis results from lattice decoding are still applicable to MIMO detection, by simply removing lattice reduction from the Markov mixing.
Additionally, besides MIMO detection, the sampler decoding strategy can also be extended to signal processing as an useful signal estimator or detector \cite{Luo1,xia2,Wuqihui1,Zhuang1,Xiangmin1}.

The rest of this paper is organized as follows. Section II introduces the background of lattice Gaussian distribution and briefly reviews the basics of Gibbs sampling as well as lattice reduction. In Section III, the convergence rate of systematic scan Gibbs sampling is derived, which is essentially determined by the HGR maximal correlation among the multivariate. Based on it, the lattice-reduction-aided Gibbs sampling algorithm is proposed in Section IV for a better Markov mixing performance. Section V extends the lattice-reduction-aided Gibbs sampling to lattice decoding. The choice of the standard deviation $\sigma$ is studied in full details while the startup mechanism of Gibbs sampling resorted to the correct decoding radius is established. Simulation results for large-scale MIMO detection are presented in Section VI. Finally, Section VII concludes the paper.

\emph{Notation:} Matrices and column vectors are denoted by upper and lowercase boldface letters, and the transpose, inverse, pseudoinverse of a matrix $\mathbf{B}$ by $\mathbf{B}^T, \mathbf{B}^{-1},$ and $\mathbf{B}^{\dag}$, respectively. We use $\mathbf{b}_i$ for the $i$th column of the matrix $\mathbf{B}$, $\mathbf{\widehat{b}}_i$ for the $i$th
Gram-Schmidt vector of the matrix $\mathbf{B}$, $b_{i,j}$ for the entry in the $i$th row and $j$th column of the matrix $\mathbf{B}$. In addition, in this paper, the computational complexity is measured by the number of arithmetic operations (additions, multiplications, comparisons, etc.). Finally, $h\in L_0^2(\pi)$ and $L_0^2(\pi)$ denote the set of all mean zero and finite variance functions with respect to the target distribution $\pi$, i.e., $E_{\pi}[h(\mathbf{x})]=0$ and $\text{var}_{\pi}[h(\mathbf{x})]=v<\infty$.

\newtheorem{my1}{Lemma}
\newtheorem{my2}{Theorem}
\newtheorem{my3}{Definition}
\newtheorem{my4}{Proposition}
\newtheorem{my5}{Corollary}
\newtheorem{my6}{Remark}

\section{Preliminaries}
In this section, we introduce the background and mathematical tools needed to describe and analyze the following lattice-reduction-aided Gibbs sampling.

\subsection{Lattice Gaussian Distribution}
Let $\mathbf{B}=[\mathbf{b}_1,\ldots,\mathbf{b}_n]\subset \mathbb{R}^n$ consist of $n$ linearly independent vectors. The $n$-dimensional lattice $\Lambda$ generated by $\mathbf{B}$ is defined by
\begin{equation}
\Lambda=\mathcal{L}(\mathbf{B})=\{\mathbf{Bx}: \mathbf{x}\in \mathbb{Z}^n\},
\end{equation}
where $\mathbf{B}$ is called the lattice basis. We define the Gaussian function centered at $\mathbf{c}\in \mathbb{R}^n$ for standard deviation $\sigma>0$ as
\begin{equation}
\rho_{\sigma, \mathbf{c}}(\mathbf{z})=e^{-\frac{\|\mathbf{z}-\mathbf{c}\|^2}{2\sigma^2}},
\end{equation}
for all $\mathbf{z}\in\mathbb{R}^n$. When $\mathbf{c}$ or $\sigma$ are not specified, we assume that they are $\mathbf{0}$ and $1$ respectively. Then, the \emph{discrete Gaussian distribution} over $\Lambda$ is defined as
\begin{equation}
D_{\Lambda,\sigma,\mathbf{c}}(\mathbf{x})=\frac{\rho_{\sigma, \mathbf{c}}(\mathbf{Bx})}{\rho_{\sigma, \mathbf{c}}(\Lambda)}=\frac{e^{-\frac{1}{2\sigma^2}\parallel \mathbf{Bx}-\mathbf{c} \parallel^2}}{\sum_{\mathbf{x} \in \mathbb{Z}^n}e^{-\frac{1}{2\sigma^2}\parallel \mathbf{Bx}-\mathbf{c} \parallel^2}}
\label{lattice gaussian distribution}
\end{equation}
for all $\mathbf{x}\in \mathbb{Z}^n$, where $\rho_{\sigma, \mathbf{c}}(\Lambda)\triangleq \sum_{\mathbf{\mathbf{Bx}}\in\Lambda}\rho_{\sigma, \mathbf{c}}(\mathbf{Bx})$ is just a scaling to obtain a probability distribution.
We remark that this definition differs slightly from the one in \cite{MicciancioGaussian}, where $\sigma$ is
scaled by a constant factor $\sqrt{2\pi}$ (i.e., $s =\sqrt{2\pi}\sigma$).
Fig. 1 illustrates the discrete Gaussian distribution over $\mathbb{Z}^2$. As can be seen clearly, it resembles a continuous Gaussian distribution, but is only defined over a lattice. In fact, discrete and continuous
Gaussian distributions share similar properties, if the \emph{flatness
factor} is small \cite{LLBS_12}.



\begin{figure}[t]
\vspace{-2em}
\hspace{-1em}\includegraphics[width=3.6in,height=2.2in]{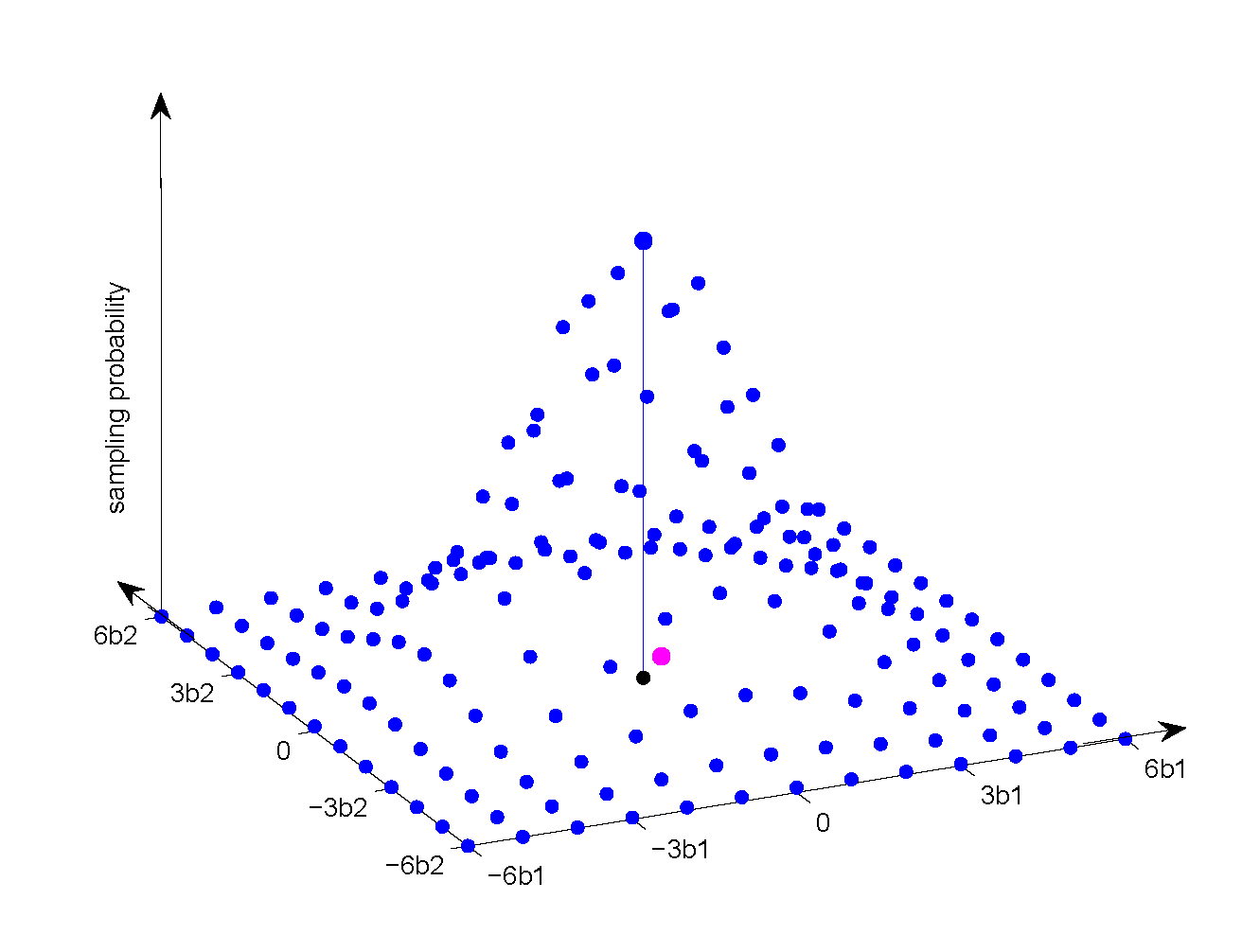}
\vspace{-2em}
  \caption{Illustration of a two-dimensional lattice Gaussian distribution with $\mathbf{B}=[\mathbf{b}_1 \mathbf{b}_2]$, where the red dot and blue dots respectively correspond to the query point $\mathbf{c}$ and sampling probabilities of candidate lattice points $D_{\Lambda,\sigma,\mathbf{c}}(\mathbf{x})$. For simplicity, only the closest lattice point $\mathbf{Bx}_{\text{cvp}}$ (black dot) with the largest sampling probability is depicted.}
  \label{simulation x}
\end{figure}

\subsection{Sampler Decoding}
Consider the decoding of an $n \times n$ real-valued system. The
extension to the complex-valued system is straightforward \cite{CongRandom,Xia1}. Let $\mathbf{x}\in \mathbb{Z}^n$ denote the transmitted signal.
The corresponding received signal $\mathbf{c}$ is given by
\begin{equation}
\mathbf{c}=\mathbf{B}\mathbf{x}+\mathbf{w}
\label{eqn:System Model}
\end{equation}
where $\mathbf{w}$ is the noise vector with zero
mean and variance $\sigma_w^{2}$, $\mathbf{B}$ is an $n\times n$ full column-rank matrix of
channel coefficients. Typically, the conventional maximum likelihood (ML) reads
\begin{equation}
\widehat{\mathbf{x}}=\underset{\mathbf{x}\in \mathcal{X}^{n}}{\operatorname{arg~min}} \, \|\mathbf{H}\mathbf{x}-\mathbf{y}\|^2
\label{eqn:ML Decoding}
\end{equation}
where $\| \cdot \|$ denotes the Euclidean norm. Clearly, ML decoding corresponds to the CVP. If the received signal $\mathbf{c}$ is the origin, then ML decoding reduces to SVP.

Intuitively, the CVP given in (\ref{eqn:ML Decoding}) can be solved by lattice Gaussian sampling. Since the distribution is centered at the query point $\mathbf{c}$, the closest lattice point $\mathbf{Bx}_{\text{cvp}}$ to $\mathbf{c}$ is assigned the largest sampling probability. Therefore, by multiple samplings, $\mathbf{x}_{\text{cvp}}$ is most likely to be returned.
It has been demonstrated that lattice Gaussian sampling is equivalent to CVP via a polynomial-time dimension-preserving reduction \cite{DGStoCVPSVP}.
More specifically, in \cite{Klein}, Klein introduced a lattice decoding algorithm which performs the sampling from a Gaussian-like distribution, and it was further improved in \cite{CongRandom,DerandomizedJ}.
Aggarwal \textit{et al.} used lattice Gaussian sampling to solve CVP and SVP with $2^n$ space and time complexities  \cite{RegevSolvingtheShortestVectorProblem} \cite{RegevSolvingtheClosestVectorProblem}.
Furthermore, only polynomial space complexity is required by the independent MHK sampling algorithm with CVP complexity $O(e^{d^2(\Lambda, \mathbf{c})/\min_i^2\|\widehat{\mathbf{b}}_i\|})$ \cite{ZhengWangTIT17}.

Decoding by sampling has promising advantages. Firstly, sampling has the potential to be efficiently implemented, thus providing a prospective decoding method especially for high-dimensional systems.
Secondly, the standard deviation $\sigma$ of the discrete Gaussian distribution can be optimized to improve the sampling probability of the target point, which leads to a better decoding performance. Thirdly, by adjusting the sample size, the sampler decoding enjoys a flexible trade-off between performance and complexity. However, the problem of sampler decoding chiefly lies on how to perform the sampling over the target lattice Gaussian distribution.

\subsection{Gibbs Sampling}
As a foremost sampling scheme in MCMC, Gibbs tries to tackles with the sampling from a complicated joint distribution through conditional sampling over its marginal distributions.
Typically, as for lattice Gaussian sampling, each coordinate of $\mathbf{x}$ is sampled from the following 1-dimensional conditional distribution
\begin{equation}
P_i(x_i|\mathbf{x}_{[-i]})\hspace{-.2em}=\hspace{-.2em}D_{\Lambda,\sigma,\mathbf{c}}(x_i|\mathbf{x}_{[-i]})\hspace{-.2em}=\hspace{-.2em}\frac{e^{-\frac{1}{2\sigma^2}\parallel \mathbf{Bx}-\mathbf{c} \parallel^2}}{\sum_{x_i\in \mathbb{Z}}e^{-\frac{1}{2\sigma^2}\parallel \mathbf{Bx}-\mathbf{c} \parallel^2}}
\label{x7z}
\end{equation}
with $\sigma>0$. Here $1\leq i\leq n$ denotes the coordinate index of $\mathbf{x}$, $\mathbf{x}_{[-i]}\triangleq[x_1,\ldots,x_{i-1},x_{i+1},\ldots,x_{n}]^T$. During this univariate sampling, the other $n-1$ variables contained in $\mathbf{x}_{[-i]}$ are leaving unchanged. By repeating such a procedure with a certain scan scheme, a Markov chain $\{\mathbf{X}^0, \mathbf{X}^1, \ldots\}$ is established.

In particular, there are various scan schemes to proceed the component updating. Apart from the random scan who randomly updates the coordinate of $\mathbf{x}$, systematic scan proceeds the update in a sequential order from $x_n$ to $x_1$, thus completing an iteration during each Markov move.
Compared to random scan, systematic scan is more preferable in lattice decoding due to its fixed update order. In fact, the mixing times of these two scan schemes do not differ by more than a polynomial factor \cite{ScanorderHe}.

\begin{my2}[\hspace{-0.005em}\cite{ITW2017}]
Given the invariant lattice Gaussian distribution $D_{\Lambda,\sigma,\mathbf{c}}$, the Markov chain induced by random scan Gibbs algorithm is geometrically ergodic
\end{my2}
\vspace{-1em}
\begin{equation}
\|P^t(\mathbf{x}, \cdot)-D_{\Lambda,\sigma,\mathbf{c}}\|_{TV}\leq C(\mathbf{x})\varrho^t
\label{geo-ergodic}
\end{equation}
\emph{with convergence rate} $\varrho<1$ and $C(\mathbf{x})<\infty$ for all $\mathbf{x}$.

Here, $C(\cdot)$ is parameterized by the initial Markov state $\mathbf{x}$ \footnote{Once $C$ is a constant for all $\mathbf{x}$, the Markov chain is referred to as uniform ergodicity. More details can be found in \cite{RobertsGeneralstatespace}.}, which also plays an important role in the Markov mixing. $\|\cdot\|_{TV}$ represents the total variation distance, $P^t(\mathbf{x}, \cdot)$ denotes the row of $\mathbf{P}^t$ corresponding to initial state $\mathbf{x}$. Another thing should be pointed out is that in MCMC the complexity of each Markov move is often insignificant, whereas the required mixing time as well as the convergence rate are more critical.

\renewcommand{\algorithmicrequire}{\textbf{Input:}}  
\renewcommand{\algorithmicensure}{\textbf{Output:}} 

\begin{algorithm}[t]
\caption{LLL Reduction}
\begin{algorithmic}[1]
\Require
$\mathbf{B}=[\mathbf{b}_1,\ldots,\mathbf{b}_n]$
\Ensure
$\overline{\mathbf{B}}=\mathbf{B}\mathbf{U}$
\State compute Gram-Schmidt orthogonality (GSO) $\widehat{\mathbf{B}}$
\State $k$=2
\While{$k\leq n$}
\State size-reduce $\mathbf{b}_k$ against $\mathbf{b}_{k-1}$
\If{$\|\widehat{\mathbf{b}}_k\|^2<(\delta-|\mu_{k,k-1}|^2)\|\widehat{\mathbf{b}}_{k-1}\|^2$}
\State swap $\mathbf{b}_k$ and $\mathbf{b}_{k-1}$ update GSO
\State $k=$max$(k-1,2)$
\Else
\For{$l=k-2, k-3,\ldots,1$}
\State size-reduce $\mathbf{b}_k$ against $\mathbf{b}_{l}$
\EndFor
\State $k=k+1$
\EndIf
\EndWhile
\State return $\overline{\mathbf{B}}=\mathbf{B}$
\end{algorithmic}
\end{algorithm}

\subsection{Lattice Reduction Technique}
Lattice reduction techniques have a long tradition in the field of number theory.
In 1982, the celebrated LLL algorithm was proposed as a powerful and famous lattice reduction criterion for arbitrary lattice.
Specifically, a basis $\mathbf{B}$ is said to be LLL-reduced\footnote{Other
lattice reduction schemes like Korkin-Zolotarev (KZ) reduction and Seysen reduction also exist, which are out of scope of this work. See \cite{WubbenLRMagzine,Shanxiang1} for more details.}, if it satisfies the following two conditions,
\begin{itemize}
  \item $|\mu_{i,j}|\leq\frac{1}{2}$,\, for \, $1\leq j<i \leq n$;

  \item $\delta\|\mathbf{\widehat{b}}_i\|^{2} \leq \|\mu_{i+1,i} \mathbf{\widehat{b}}_i + \mathbf{\widehat{b}}_{i+1}\|^{2}$,\, for \, $1\leq i<n$.
\end{itemize}
The first clause is called size reduction condition with $\mu_{i,j}=\langle\mathbf{b}_i,\mathbf{\widehat{b}}_j\rangle/\langle\mathbf{\widehat{b}}_j,\mathbf{\widehat{b}}_j\rangle$, while the second is known as Lov\'{a}sz condition. If Lov\'{a}sz condition is violated, the basis vectors $\mathbf{b}_i$ and $\mathbf{b}_{i+1}$ are swapped; otherwise, size reduction is carried out.
If only size reduction condition is satisfied, then the basis is called size-reduced.
The parameter $1/4 < \delta <1$ controls both the convergence speed of the reduction and the degree of orthogonality of the reduced basis \cite{Adeane}.

After LLL reduced, the lattice basis consists of vectors that are relatively short and orthogonal to each other. More precisely, LLL reduction is able to yield a lattice vector within $(2/\sqrt{3})^n$ of the shortest vector in lattice by average polynomial complexity $O(n^4\text{log}n)$ \cite{LLLoriginal}. Inspired by it, the lattice-reduction-aided decoding has emerged as a powerful decoding strategy in various research fields. In MIMO detection, it has been demonstrated that the LLL reduction based minimum mean square error (MMSE) detection not only attains the full receive diversity \cite{MTaherzadeh2007}, but also facilitates the diversity-multiplexing trade-off (DMT) optimal decoding \cite{JaldenDMT}. Meanwhile, LLL reduction can be efficiently realized by \emph{effective LLL reduction} with polynomial complexity $O(n^3\text{log}\,n)$ \cite{CongEffective}. Nevertheless, the performance gap between the optimal ML decoding and lattice-reduction-aided decoding is still substantial especially in high-dimensional systems.

%

\subsection{HGR Maximal Correlation}
For decades, the measurement of Hirschfeld-Gebelein-R\'enyi (HGR) maximal correlation has found numerous interesting applications in the field of information theory \cite{MaximalCorrelation2,MaximalCorrelation3,MaximalCorrelation4}.
\begin{my3}[\hspace{-0.005em}\cite{MaximalCorrelation1}]
For any two random variables $\xi$ and $\eta$, their maximal correlation $\gamma(\xi, \eta)$ is defined as
\end{my3}
\vspace{-1em}
\begin{equation}
\gamma(\xi, \eta)=\underset{f,g}{\sup}\ \text{corr}(f(\xi), g(\eta)),
\end{equation}
\emph{where the supremum is taken over all Borel functions with }$0<\text{var}[f(\xi)]<\infty$ \emph{and} $0<\text{var}[g(\eta)]<\infty$.

More specifically, with $\text{E}[f(\xi)]=\text{E}[g(\eta)]=0$ and $\text{var}[f(\xi)]=\text{var}[g(\eta)]=1$, the HGR maximal correlation can be rewritten as
\begin{equation}
\gamma(\xi, \eta)=\underset{f,g}{\sup}\ \text{E}[f(\xi), g(\eta)].
\end{equation}
Then, by a simple application of the \emph{Cauchy-Schwarz inequality}, R\'enyi showed the following one-function alternate characterization for $\gamma(\xi, \eta)$ as \cite{MaximalCorrelation4},
\begin{equation}
\gamma^2(\xi, \eta)=\underset{f(\xi):\text{E}(f)=0, \text{var}(f)=1}{\sup}\text{E}[\text{E}^2[f(\xi)|\eta]].
\end{equation}

Theoretically, HGR maximal correlation is an elegant generalization of the well-known \emph{Pearson correlation coefficient}, and serves as a normalized measure of the dependence between two random variables. Although the Pearson correlation is analytically simple to evaluate in theory and computationally tractable to implement in practice, it only measures the linear relationship between $\xi$ and $\eta$ rather than capturing true statistical dependence. Apart from Pearson correlation coefficient, $\gamma(\xi, \eta)$ is defined whenever both $\xi$ and $\eta$ are non-degenerate, which assumes values in the interval $[0,1]$ and vanish if and only if $\xi$ and $\eta$ are independent.

\section{Convergence Analysis}
In this section, the convergence analysis of systematic scan Gibbs sampling for lattice Gaussian sampling is presented, where its convergence rate is derived by means of HGR maximal correlation.

\subsection{Systematic Scan Gibbs Sampling}
To start with, by induction, the transition probability of the systematic scan Gibbs sampling can be expressed as
\begin{equation}
P(\mathbf{X}^t\hspace{-.2em}=\hspace{-.2em}\mathbf{x},\hspace{-.2em}\mathbf{X}^{t+1}\hspace{-.4em}=\hspace{-.3em}\mathbf{y})\hspace{-.3em}=\hspace{-.2em}\prod_{i=1}^n\hspace{-.3em}P_{n-i+1}(x^{t+1}_{n-i+1}|\mathbf{x}^t_{[-(n-i+1)]}).
\label{eq:transition2}
\end{equation}
Clearly, for a given standard deviation $\sigma>0$ and full rank lattice basis $\mathbf{B}$, it is easy to verify that each random variable $x_i$ is sampled with variance
\begin{equation}
\text{var}[x_i|\mathbf{x}_{[-i]}]=\kappa_i>0.
\label{correlationsetup}
\end{equation}
Therefore, all the sampling candidates of $x_i$ are possible to be sampled theoretically, indicating an irreducible chain. In principle, the irreducible property prevents the random variables to be totally dependent, where all the components of $\mathbf{x}$ for Markov state $\mathbf{X}^t$ may be different with $\mathbf{y}$ of $\mathbf{X}^{t+1}$.

For the sake of convergence analysis, we formulate the systematic scan Gibbs sampling to a simple version which only consists of two nominal components $\mathbf{x}=[\mathbf{x}_1; \mathbf{x}_2]$, $\mathbf{x}_1\in\mathbb{Z}^m$ and $\mathbf{x}_2\in\mathbb{Z}^{n-m}$. In particular, similar to (\ref{x7z}), during a Markov move, $\mathbf{x}_1$ and $\mathbf{x}_2$ are iteratively generated by
\begin{eqnarray}
\mathbf{x}^{t+1}_2\sim P_{\mathbf{x}_2}(\mathbf{x}_2|\mathbf{x}^t_1)=\frac{e^{-\frac{1}{2\sigma^2}\parallel \mathbf{Bx}-\mathbf{c} \parallel^2}}{\sum_{\mathbf{x}_2\in \mathbb{Z}^{n-m}}e^{-\frac{1}{2\sigma^2}\parallel \mathbf{Bx}-\mathbf{c} \parallel^2}}
\label{x78}
\end{eqnarray}
and
\begin{eqnarray}
\mathbf{x}^{t+1}_1\sim P_{\mathbf{x}_1}(\mathbf{x}_1|\mathbf{x}^{t+1}_2)=\frac{e^{-\frac{1}{2\sigma^2}\parallel \mathbf{Bx}-\mathbf{c} \parallel^2}}{\sum_{\mathbf{x}_1\in \mathbb{Z}^m}e^{-\frac{1}{2\sigma^2}\parallel \mathbf{Bx}-\mathbf{c} \parallel^2}}.
\label{x77}
\end{eqnarray}


In contrast to the conventional \emph{data augmentation} scheme in MCMC, sampling over subvectors $\mathbf{x}_1$ and $\mathbf{x}_2$ can be conducted via blocked sampling \cite{RobertsAndSahu}, which does enable a faster convergence by taking multiple sampling elements into account (see \cite{ZhengWangMCMCLatticeGaussian} for an efficient blocked strategy of Gibbs sampler for lattice Gaussian sampling). We also claim that the following convergence analysis with respect to such a simplification can be easily adopted to general cases with $m=1$ (e.g., $\mathbf{x}_1\in\mathbb{Z}^1$ and $\mathbf{x}_2\in\mathbb{Z}^{n-1}$).

Through the simplification, the above Markov chain still attains $\pi=D_{\Lambda,\sigma,\mathbf{c}}$ as the invariant distribution while its transition probability becomes
\begin{equation}
P(\mathbf{X}^t\hspace{-.2em}=\hspace{-.2em}\mathbf{x},\hspace{-.2em} \mathbf{X}^{t+1}\hspace{-.2em}=\hspace{-.2em}\mathbf{y}\hspace{-.1em})\hspace{-.2em}=\hspace{-.2em}P_{\mathbf{x}_{1}}(\mathbf{x}^{t+1}_{1}|\mathbf{x}^t_{2})\hspace{-.2em}\cdot\hspace{-.2em} P_{\mathbf{x}_{2}}(\mathbf{x}^{t+1}_{2}|\mathbf{x}^{t+1}_{1}).
\label{eq:transition12}
\end{equation}
Insight into this simplified Gibbs sampler, the marginal chains $\{\mathbf{x}^1_1, \mathbf{x}^2_1, \ldots\}$ and $\{\mathbf{x}^1_2, \mathbf{x}^2_2, \ldots\}$ with respect to $\mathbf{x}_1$ and $\mathbf{x}_2$ also function as valid Markov chains. Most importantly, these marginal chains experience the same mixing performance as the original chain with convergence rate \cite{YosidaBook,LiuCovarianceSchemes}
\begin{equation}
\varrho=\varrho_1=\varrho_2,
\label{x12}
\end{equation}
which implies we can obtain the convergence rate of the joint chain by only focusing on its marginal chain.
Furthermore, because $\mathbf{x}^{t}_1$ and $\mathbf{x}^{t+1}_1$ are conditionally independent for a given $\mathbf{x}^{t+1}_2$,
the \emph{detailed balance condition} is satisfied by
\begin{equation*}
\pi'(\mathbf{x}^t_1)P(\mathbf{x}^t_1,\mathbf{x}^{t+1}_1)\hspace{-.3em}=\pi'(\mathbf{x}^t_1)\sum_{\mathbf{x}^{t+1}_2}\pi(\mathbf{x}^{t+1}_2|\mathbf{x}^t_1)\pi(\mathbf{x}^{t+1}_1|\mathbf{x}^{t+1}_2)
\end{equation*}
\vspace{-1em}
{\allowdisplaybreaks\begin{flalign}
\hspace{-.3em}=&\sum_{\mathbf{x}^{t+1}_2}\hspace{-.2em}\pi(\mathbf{x}^t_1|\mathbf{x}^{t+1}_2)\pi(\mathbf{x}^{t+1}_2|\mathbf{x}^t_1)\pi(\mathbf{x}^{t+1}_1|\mathbf{x}^{t+1}_2)\notag\\
\hspace{-.3em}=&\hspace{-.5em}\sum_{\mathbf{x}^{t+1}_2}\hspace{-.2em}\pi(\mathbf{x}^{t+1}_1|\mathbf{x}^{t+1}_2)\pi(\mathbf{x}^{t+1}_2|\mathbf{x}^{t+1}_1)\pi(\mathbf{x}^t_1|\mathbf{x}^{t+1}_2)\notag\\
\hspace{-.3em}=&\pi'(\mathbf{x}^{t+1}_1)\sum_{\mathbf{x}^{t+1}_2}\pi(\mathbf{x}^{t+1}_2|\mathbf{x}^{t+1}_1)\pi(\mathbf{x}^t_1|\mathbf{x}^{t+1}_2)\notag\\
\hspace{-.3em}=&\pi'(\mathbf{x}^{t+1}_1)P(\mathbf{x}^{t+1}_1,\mathbf{x}^t_1),
\end{flalign}}indicating that the marginal chain turns out to be reversible. Inspired by it, the following convergence analysis takes place in the marginal Markov chain $\{\mathbf{x}^1_1, \mathbf{x}^2_1, \ldots\}$ with target distribution $\pi'$ for simplicity\footnote{The same result can be obtained with respect to the marginal Markov chain $\{\mathbf{x}^1_2, \mathbf{x}^2_2, \ldots\}$.}.


\subsection{Convergence Analysis}
Typically, given the transition probability $P(\mathbf{X}^t, \mathbf{X}^{t+1})$, the forward operator $\mathbf{F}$ of the Markov chain is defined as \cite{LiuBook}
\begin{equation}
\mathbf{F}h(\mathbf{X}^t)\triangleq\hspace{-.8em}\sum_{\mathbf{X}^{t+1}\in\Omega}\hspace{-1em}h(\mathbf{X}^{t+1})P(\mathbf{X}^t, \mathbf{X}^{t+1})\hspace{-.2em}=\hspace{-.2em}\text{E}[h(\mathbf{X}^{t+1})|\mathbf{X}^t]
\label{xp}
\end{equation}
with induced operator norm
\begin{equation}
\|\mathbf{F}\|=\hspace{-1em}\underset{h\in L_0^2(\pi), \text{var}(h)=1}{\sup}\hspace{-1em}\|\mathbf{F}h\|.
\label{xp1}
\end{equation}
Here, $L^2(\pi)$ is the Hilbert space of square integrable functions with respect to $\pi$ so that $L_0^2(\pi)\triangleq\{h(\mathbf{x}):\text{E}[h(\mathbf{x})]=0, \text{var}[h(\mathbf{x})]<\infty\}$ denotes the subspace of $L^2(\pi)$ consisting of functions with zero mean relative to $\pi$. More precisely, for $h(\cdot),g(\cdot)\in L_0^2(\pi)$, the inner product defined by the space is
\begin{equation}
\langle h(\mathbf{x}),g(\mathbf{x})\rangle=\text{E}[h(\mathbf{x})g(\mathbf{x})]
\end{equation}
with variance
\begin{equation}
\text{var}_{\pi}[h(\mathbf{x})]=\langle h(\mathbf{x}), h(\mathbf{x})\rangle=\|h(\mathbf{x})\|^2.
\label{m3}
\end{equation}

\begin{my2}
Given the invariant lattice Gaussian distribution $\pi=D_{\Lambda,\sigma,\mathbf{c}}$, the Markov chain induced by systematic scan Gibbs algorithm is geometrically ergodic
\end{my2}
\vspace{-1em}
\begin{equation}
\|P^t(\mathbf{x}, \cdot)-D_{\Lambda,\sigma,\mathbf{c}}\|_{TV}\leq C(\mathbf{x})\varrho^t
\label{geo-ergodic}
\end{equation}
\emph{with convergence rate}
\begin{equation}
\varrho=\gamma^2(\mathbf{x}_1, \mathbf{x}_2)<1.
\end{equation}
\begin{proof}
First of all, regarding to the marginal Markov chain $\{\mathbf{x}^1_1, \mathbf{x}^2_1, \ldots\}$, the spectral radius of $\mathbf{F}_1$ is closely related with its norm as \cite{Fill1991}
\begin{equation}
\text{spec}(\mathbf{F}_1)=\underset{t\rightarrow\infty}{\lim}\|\mathbf{F}_1^t\|^{1/t}.
\label{x9}
\end{equation}

\begin{figure*}[t]
\centering\includegraphics[width=6.5in]{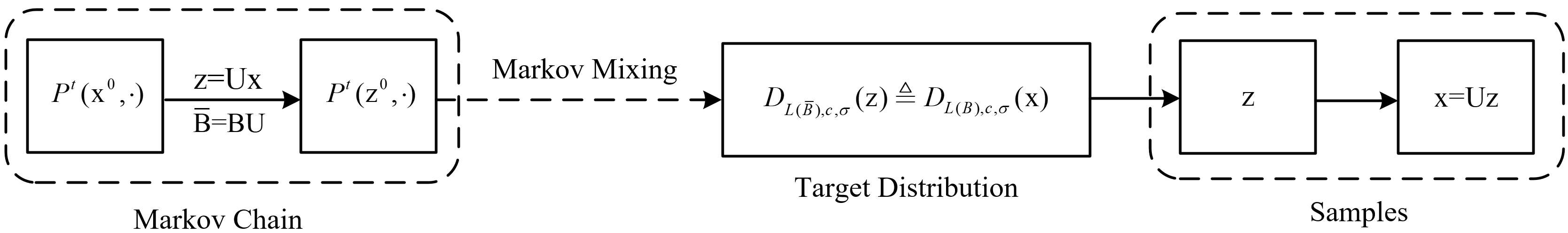}
\caption{Illustration of the lattice-reduction-aided Gibbs sampler for lattice Gaussian sampling.}
\label{IDD figure}
\end{figure*}

Meanwhile, the reversibility of the marginal chain corresponds to a self-adjoint operator $\mathbf{F}_1$ with \cite{LiuCovarianceScans}
\begin{equation}
\|\mathbf{F}_1^t\|=\|\mathbf{F}_1\|^t,
\end{equation}
then we have
\begin{equation}
\text{spec}(\mathbf{F}_1)=\|\mathbf{F}_1\|.
\label{m2}
\end{equation}
Subsequently, according to (\ref{xp1}) and (\ref{m2}), the spectral radius of the forward operator $\mathbf{F}_1$ is derived as
{\allowdisplaybreaks\begin{flalign}
\text{spec}(\mathbf{F}_1)&=\|\mathbf{F}_1\|=\underset{h\in L_0^2(\pi'), \text{var}(h)=1}{\sup}\hspace{-1em}\|\mathbf{F}_1h\|\notag\\
&=\hspace{-1em}\underset{h\in L_0^2(\pi'), \text{var}(h)=1}{\sup}\hspace{-1em}\{\text{var}[\text{E}[h(\mathbf{x}_1^{t+1})|\mathbf{x}_1^t]]\}^{\frac{1}{2}}\notag\\
&=\hspace{-2.5em}\underset{h\in L_0^2(\pi'), \text{var}(h)=1}{\sup}\hspace{-2.5em}\{\text{E}[\text{E}^2[h(\mathbf{x}_1^{t+1})|\mathbf{x}_1^t]]-[\text{E}[\text{E}[h(\mathbf{x}_1^{t+1})|\mathbf{x}_1^t]]]^2\}^{\frac{1}{2}}\notag\\
&=\hspace{-1em}\underset{h\in L_0^2(\pi'), \text{var}(h)=1}{\sup}\hspace{-1em}\{\text{E}[\text{E}^2[h(\mathbf{x}_1^{t+1})|\mathbf{x}_1^t]]\}^{\frac{1}{2}}\notag\\
&=\gamma(\mathbf{x}^t_1,\mathbf{x}^{t+1}_1).
\label{x11}
\end{flalign}}With respect to $\gamma(\mathbf{x}^t_1,\mathbf{x}^{t+1}_1)$, on one hand, it follows that
{\allowdisplaybreaks\begin{flalign}
\gamma(\mathbf{x}^t_1,\mathbf{x}^{t+1}_1)&=\hspace{-1em}\underset{h\in L_0^2(\pi'), \text{var}(h)=1}{\sup}\hspace{-1em}\text{var}[\text{E}[\text{E}[h(\mathbf{x}_1^{t+1})|\mathbf{x}_2^{t+1}]|\mathbf{x}_1^t]]\notag\\
&\leq\hspace{-1em}\underset{h\in L_0^2(\pi'), \text{var}(h)=1}{\sup}\hspace{-1em}\text{var}[\text{E}[h(\mathbf{x}_1^{t+1})|\mathbf{x}_2^{t+1}]]\notag\\
&=\hspace{-1em}\underset{h\in L_0^2(\pi'), \text{var}(h)=1}{\sup}\hspace{-1em}\text{E}[\text{E}^2[h(\mathbf{x}_1)\mid\mathbf{x}_2]]\notag\\
&=\gamma^2(\mathbf{x}_1,\mathbf{x}_2).
\label{lll1}
\end{flalign}}On the other hand, we have
{\allowdisplaybreaks\begin{flalign}
\gamma(\mathbf{x}^t_1,\mathbf{x}^{t+1}_1)&\geq\hspace{-1em}\underset{h\in L_0^2(\pi'), \text{var}(h)=1}{\sup}\hspace{-1em}\text{corr}[h(\mathbf{x}_1^t),h(\mathbf{x}_1^{t+1})]\notag\\
&=\hspace{-1em}\underset{h\in L_0^2(\pi'), \text{var}(h)=1}{\sup}\hspace{-1em}\text{E}[h(\mathbf{x}_1^t)h(\mathbf{x}_1^{t+1})]\notag\\
&=\hspace{-1em}\underset{h\in L_0^2(\pi'), \text{var}(h)=1}{\sup}\hspace{-1em}\text{E}[\text{E}[h(\mathbf{x}_1^t)h(\mathbf{x}_1^{t+1})\mid\mathbf{x}_2^{t+1}]]\notag\\
&=\hspace{-1em}\underset{h\in L_0^2(\pi'), \text{var}(h)=1}{\sup}\hspace{-1em}\text{E}[\text{E}^2[h(\mathbf{x}_1)\mid\mathbf{x}_2]]\notag\\
&=\gamma^2(\mathbf{x}_1,\mathbf{x}_2).
\label{lll2}
\end{flalign}}Therefore, according to (\ref{lll1}) and (\ref{lll2}), we get
\begin{equation}
\text{spec}(\mathbf{F}_1)=\gamma(\mathbf{x}^t_1,\mathbf{x}^{t+1}_1)=\gamma^2(\mathbf{x}_1,\mathbf{x}_2)<1,
\end{equation}
where the inequality holds due to the fact that $\mathbf{x}_1$ and $\mathbf{x}_2$ are random variables of each other by configuration.

Next, by invoking the following Lemma from \cite{SpectralGapL2}, the marginal chain $\{\mathbf{x}^1_1, \mathbf{x}^2_1, \ldots\}$ turns out to be geometrically ergodic with convergence rate
\begin{equation}
\varrho_1=\text{spec}(\mathbf{F}_1).
\label{mng1}
\end{equation}


\begin{my1}[\hspace{-0.005em}\cite{SpectralGapL2}]
Given the invariant distribution $\pi$, a reversible, irreducible and aperiodic Markov chain with spectral gap $\gamma=1-\text{spec}(\mathbf{F})>0$ converges exponentially as
\end{my1}
\vspace{-1em}
\begin{equation}
\|P^t(\mathbf{x}, \cdot)-\pi(\cdot)\|_{TV}\leq C(\mathbf{x})(1-\gamma)^t.
\label{m1}
\end{equation}

Hence, from (\ref{x12}) and (\ref{mng1}), the original Markov chain $\{\mathbf{X}^1, \mathbf{X}^2, \ldots\}$ is geometric ergodicity with exponential convergence rate
\begin{equation}
\varrho=\gamma^2(\mathbf{x}_1,\mathbf{x}_2)<1,
\end{equation}
completing the proof.
\end{proof}

Clearly, $0\leq\gamma(\mathbf{x}_1,\mathbf{x}_2)<1$ measures the dependence between $\mathbf{x}_1$ and $\mathbf{x}_2$, where $\gamma(\mathbf{x}_1,\mathbf{x}_2)=0$ if and only if $\mathbf{x}_1$ and $\mathbf{x}_2$ are independent of each other. On the other hand, the high correlation between $\mathbf{x}_1$ and $\mathbf{x}_2$ gives rise to a larger value of $\gamma(\mathbf{x}_1,\mathbf{x}_2)$ approaching to $1$.
It should be noticed that such a result can be easily generalized as $\varrho=\gamma^2(x_i,\mathbf{x}_{[-i]})$, where a less correlation among $x_i$ and $\mathbf{x}_{[-i]}$ for $1\leq i\leq n$ is also the sufficient condition for a small value of $\gamma(\mathbf{x}_1,\mathbf{x}_2)$.

\begin{my6}
The convergence rate of systematic scan Gibbs sampling for the lattice Gaussian distribution $D_{\Lambda,\sigma,\mathbf{c}}$ is dominated by the HGR maximal correlation $\gamma(x_i,\mathbf{x}_{[-i]})$ among random variables $x_i$'s, where the optimal convergence $\varrho=0$ happens when $x_i$'s are independent of each other.
\end{my6}

\section{Lattice-reduction-aided Gibbs Sampling}
From the convergence analysis, in order to achieve an efficient Markov mixing, a smaller $\gamma(x_i,\mathbf{x}_{[-i]})$, $1\leq i\leq n$ is strongly desired. However, it is hard to explicitly calculate $\gamma$ in practice. Regarding to the lattice Gaussian distribution shown in (\ref{lattice gaussian distribution}), it is clear that the correlation over elements of $\mathbf{x}$ is decided by matrix $\mathbf{B}$, i.e., the more orthogonal of $\mathbf{B}$, the less correlation of components in $\mathbf{x}$. For this reason, we attempt to use the \emph{orthogonality defect} of $\mathbf{B}$ to partially characterize $\gamma(x_i,\mathbf{x}_{[-i]})$.

Specifically, the orthogonality defect of a matrix $\mathbf{B}$ is defined as \cite{LLLoriginal}
\begin{equation}
\xi(\mathbf{B})=\frac{\prod^n_{i=1}\|\mathbf{b}_i\|}{|\det(\mathbf{B})|},
\end{equation}
where $\det(\cdot)$ represent the determinant of the square matrix. According to \emph{Hadamard inequality}, the orthogonality defect is lower bounded by $\xi(\mathbf{B})\geq1$, where the equality holds if and only if vectors in $\mathbf{B}$ are mutually orthogonal. Consequently, we can easily arrive at the following Lemma, whose proof is omitted here due to simplicity.
\begin{my1}
If the full rank matrix $\mathbf{B}\in\mathbb{R}^{n\times n}$ is an orthogonal matrix with $\xi(\mathbf{B})=1$,
then $\gamma(x_i,\mathbf{x}_{[-i]})=0$ for $1\leq i\leq n$, and samples from lattice Gaussian distribution $D_{\Lambda,\sigma,\mathbf{c}}$ can be immediately obtained by systematic scan Gibbs sampling
with convergence rate
\end{my1}
\vspace{-1.5em}
\begin{equation}
\varrho=0.
\end{equation}

Clearly, a smaller value of $\xi(\mathbf{B})$ is in high demand for the fast mixing.
However, for a given lattice basis $\mathbf{B}$, any attempt to reduce
$\xi(\mathbf{B})$ directly for a small $\gamma(x_i,\mathbf{x}_{[-i]})$ is impossible. Nevertheless, an alternative way can still be carried out by resorting to lattice reduction technique \cite{LLLoriginal}, which transfers the lattice Gaussian distribution in (\ref{lattice gaussian distribution}) to an equivalent one:
\begin{equation}
\pi(\mathbf{z})=\frac{e^{-\frac{1}{2\sigma^2}\parallel \overline{\mathbf{B}}\mathbf{z}-\mathbf{c} \parallel^2}}{\sum_{\mathbf{z} \in \mathbb{Z}^n}e^{-\frac{1}{2\sigma^2}\parallel \overline{\mathbf{B}}\mathbf{z}-\mathbf{c} \parallel^2}},
\label{bc}
\end{equation}
where $\overline{\mathbf{B}}=\mathbf{BU}$, $\mathbf{z}=\mathbf{U}^{-1}\mathbf{x}\in\mathbb{Z}^n$ and $\mathbf{U}\in \mathbb{Z}^{n\times n}$ is a unimodular matrix with $\text{det}(\mathbf{U})=\pm1$.

Undoubtedly, $\overline{\mathbf{B}}\mathbf{z}$ and $\mathbf{B}\mathbf{x}$ describe the same lattice point in the space. Therefore, the target distribution $\pi=D_{\Lambda,\sigma,\mathbf{c}}$ essentially maintains unchanged during this transformation but is parameterized by $\mathbf{z}$, where there is a one-to-one correspondence between $\mathbf{x}$ and $\mathbf{z}$. Then, with respect to the Gibbs sampling, the conditional sampling probability of Gibbs sampling shown in (\ref{x77}) becomes
\begin{equation}
P_{\mathbf{z}_1}(\mathbf{z}_1|\mathbf{z}_2)=\frac{e^{-\frac{1}{2\sigma^2}\parallel \mathbf{\overline{B}z}-\mathbf{c} \parallel^2}}{\sum_{\mathbf{z}_1\in \mathbb{Z}^m}e^{-\frac{1}{2\sigma^2}\parallel \mathbf{\overline{B}z}-\mathbf{c} \parallel^2}}
\label{x79}
\end{equation}
with $\mathbf{z}_1\in\mathbb{Z}^m$ and $\mathbf{z}_2\in\mathbb{Z}^{n-m}$, and can be further generalized to
\begin{equation}
P_i(z_i|\mathbf{z}_{[-i]})=\frac{e^{-\frac{1}{2\sigma^2}\parallel \overline{\mathbf{B}}\mathbf{z}-\mathbf{c} \parallel^2}}{\sum_{z_i\in \mathbb{Z}}e^{-\frac{1}{2\sigma^2}\parallel \overline{\mathbf{B}}\mathbf{z}-\mathbf{c} \parallel^2}}.
\label{xv7}
\end{equation}

In particular, as shown in Fig. 2, given the target distribution $\pi(\mathbf{x})=D_{\Lambda,\sigma,\mathbf{c}}(\mathbf{x})$, the proposed lattice-reduction-aided Gibbs sampling consists of the following three steps:

1)\ \hspace{-.2em}\emph{Generate the equivalent lattice Gaussian distribution $D_{\mathcal{L}(\overline{\mathbf{B}}),\sigma,\mathbf{c}}(\mathbf{z})$ by LLL reduction.}

2)\ \emph{Perform the Gibbs sampling over $D_{\mathcal{L}(\overline{\mathbf{B}}),\sigma,\mathbf{c}}(\mathbf{z})$}.

3)\ \emph{Collect samples of $\mathbf{z}$ after the Markov mixing and output samples of $\mathbf{x}$ by $\widehat{\mathbf{x}}=\mathbf{U}\widehat{\mathbf{z}}$}.

Meanwhile, similarly, it is straightforward to verify the Gibbs sampling with respect to the converted lattice Gaussian distribution is also geometrically ergodic.
\begin{my2}
Given $D_{\mathcal{L}(\overline{\mathbf{B}}),\sigma,\mathbf{c}}(\mathbf{z})$, the Markov chain induced by Gibbs sampling converges exponentially fast:
\begin{equation}
\|P^t(\mathbf{z}, \cdot)-D_{\mathcal{L}(\overline{\mathbf{B}}),\sigma,\mathbf{c}}(\cdot)\|_{TV}\leq C'(\mathbf{z})(\varrho')^t,
\label{eq:b65}
\end{equation}
where $D_{\mathcal{L}(\overline{\mathbf{B}}),\sigma,\mathbf{c}}(\mathbf{z})\triangleq D_{\Lambda,\sigma,\mathbf{c}}(\mathbf{x})$.
\end{my2}

\begin{algorithm}[t]
\caption{Lattice-Reduction-Aided Gibbs Algorithm for Lattice Gaussian Sampling}
\begin{algorithmic}[1]
\Require
$\mathbf{B}, \sigma, \mathbf{c}, \mathbf{X}^0$, $t_{\text{mix}}(\epsilon)$
\Ensure
$\mathbf{x} \thicksim D_{\Lambda,\sigma,\mathbf{c}}$
\State let $\mathbf{x}^0$ denote the intial state of $\mathbf{X}^{0}$
\State obtain $\overline{\mathbf{B}}=\mathbf{BU}$ and $\mathbf{z}^0=\mathbf{U}^{-1}\mathbf{x}^0$ via LLL reduction
\For {$t=$1,2,\ \ldots}
\For {$i=$n,\ \ldots,\ $1$}
\State sample $z^t_i$ from $P(z_i|\mathbf{z}_{[-i]})$ shown in (\ref{xv7})
\EndFor
\State update $\mathbf{z}$ with the sampled $z_i$ and let $\mathbf{Z}^t=\mathbf{z}$
\If {$t\geq t_{\text{mix}}(\epsilon)$}
\State output the state of $\mathbf{X}^t=\mathbf{U}\mathbf{Z}^t$
\EndIf
\EndFor
\end{algorithmic}
\end{algorithm}


Remarkably, such a slight change by replacing $\mathbf{x}$ with $\mathbf{z}$ introduces a significant benefit: compared to $\mathbf{B}$, the orthogonality of matrix $\overline{\mathbf{B}}$ is greatly improved by lattice reduction.
More specifically, it has been demonstrated that after LLL reduction, the orthogonality defect of the reduced basis $\overline{\mathbf{B}}$ is upper bounded by \cite{LLLoriginal}
\begin{equation}
\xi(\overline{\mathbf{B}})\leq\beta^{\frac{n(n-1)}{4}}
\end{equation}
with $\beta=(\delta-\frac{1}{4})^{-1}$, indicating a guaranteed reduction from $\xi(\mathbf{B})$ to $\xi(\overline{\mathbf{B}})$.
Therefore, a smaller HGR maximal correlation over components within $\mathbf{z}$ is most likely to be achieved, i.e., $\gamma(\mathbf{z}_1,\mathbf{z}_2)\leq\gamma(\mathbf{x}_1,\mathbf{x}_2)$,
thus leading to a better convergence rate by Theorem 2.

\begin{my6}
With respect to sampling from the lattice Gaussian distribution $D_{\Lambda,\sigma,\mathbf{c}}$, the usage of lattice reduction is capable of achieving less correlated random variable $z_i$'s than $x_i$'s, which leads to a more efficient Markov mixing.
\end{my6}

To summarize, the proposed lattice-reduced-aided Gibbs sampling algorithm is presented in Algorithm 2.

\section{Lattice-Reduction-Aided Gibbs Sampling Algorithm for Lattice Decoding}
In this section, we extend the proposed lattice-reduction-aided Gibbs sampling to lattice decoding. Theoretically, when MCMC method is applied for sampler decoding, its decoding performance can be evaluated by CVP decoding complexity (i.e., the number of Markov move $t$), which is defined by \cite{ZhengWangTIT17}
\begin{equation}
C_{\mathrm{cvp}}\triangleq\frac{t_{\text{mix}}}{D_{\Lambda,\sigma,\mathbf{c}}(\mathbf{x}_{\text{cvp}})}.
\label{complexdf}
\end{equation}
Here, the mixing time $t_{\text{mix}}$ serves as a pick-up gap to guarantee i.i.d. samples because samples from the stationary distribution tend to be correlated with each other.
Besides, $D_{\Lambda,\sigma,\mathbf{c}}(\mathbf{x}_{\text{cvp}})$ denotes the sampling probability of the target CVP point.
Therefore, in order to strengthen the decoding performance, one can either reduce the mixing time $t_{\text{mix}}$ (e.g., use LLL reduction to boost the convergence), or improve the sampling probability $D_{\Lambda,\sigma,\mathbf{c}}(\mathbf{x}_{\text{cvp}})$, which will be studied in the following.

\subsection{Choice of the Sampling Deviation $\sigma$}
From the point of view of simulated annealing in statistics, $\sigma$ functions as ``temperature" to guide the Markov mixing, which also has an impact upon $t_{\text{mix}}$ as well.
Given the lattice Gaussian distribution $\pi(\mathbf{z})$ shown in (\ref{bc}), although a small size $\sigma$ corresponds to a relatively large decoding sampling probability $D_{\Lambda,\sigma,\mathbf{c}}(\mathbf{z}_{\text{cvp}})$, it also incurs a ``cold" Markov chain which tends to be trapped by the frozen status, and vice versa\cite{ParallelTempering2}\footnote{Actually, this is in accordance with the result of independent MHK sampling algorithm for lattice Gaussian distribution, where the exact convergence rate as well as the mixing time $t_{\text{mix}}$ can be estimated \cite{ZhengWangTIT15,ZhengWangTIT17}.}.
However, since $t_{\text{mix}}$ for Gibbs sampling is hard to get at the current stage, to balance this inherent trade-off for a better decoding performance, a feasible compromise is to ensure a reliable sampling probability given moderate $\sigma$.


In particular, with respect to any $\mathbf{z}\in\mathbb{Z}^n$ to be sampled, we firstly extract $\sigma$ from the denominator of $\pi(\mathbf{z})$ as
{\allowdisplaybreaks\begin{flalign}
\pi(\mathbf{z})&=\frac{e^{-\frac{1}{2\sigma^2}\parallel \overline{\mathbf{B}}\mathbf{z}-\mathbf{c} \parallel^2}}{\sum_{\mathbf{z} \in \mathbb{Z}^n}e^{-\frac{1}{2\sigma^2}\parallel \overline{\mathbf{B}}\mathbf{z}-\mathbf{c} \parallel^2}}\notag\\
&\overset{(a)}{\geq}\frac{e^{-\frac{1}{2\sigma^2}\parallel \overline{\mathbf{B}}\mathbf{z}-\mathbf{c} \parallel^2}}{\sum_{\mathbf{z} \in \mathbb{Z}^n}e^{-\frac{1}{2\sigma^2}\parallel \overline{\mathbf{B}}\mathbf{z}\parallel^2}}\notag\\
&\overset{(b)}{\geq}\frac{e^{-\frac{1}{2\sigma^2}\parallel \overline{\mathbf{B}}\mathbf{z}-\mathbf{c} \parallel^2}}{(\sqrt{2\pi}\sigma)^n\sum_{\mathbf{z} \in \mathbb{Z}^n}e^{-\pi\parallel \overline{\mathbf{B}}\mathbf{z}\parallel^2}}\notag\\
&=f(\sigma)\cdot c\ \ \text{for}\ \sqrt{2\pi}\sigma\geq1
\label{bc1}
\end{flalign}}where
\begin{equation}
c\triangleq1/\sum_{\mathbf{z} \in \mathbb{Z}^n}e^{-\pi\parallel \overline{\mathbf{B}}\mathbf{z}\parallel^2}
\end{equation}
is a constant and
\begin{equation}
f(\sigma)\triangleq\frac{e^{-\frac{1}{2\sigma^2}\parallel \overline{\mathbf{B}}\mathbf{z}-\mathbf{c} \parallel^2}}{(\sqrt{2\pi}\sigma)^n}
\end{equation}
is parameterized by $\sigma$. Here, $(a)$ and $(b)$ respectively obey the facts from lattice theory (\cite[Lemma 1.4]{Banaszczyk}) that
\begin{equation}
\sum_{\mathbf{v}\in\Lambda}e^{-\frac{1}{2\sigma^2}\|\mathbf{v}-\mathbf{c}\|^2}\leq\sum_{\mathbf{v}\in\Lambda}e^{-\frac{1}{2\sigma^2}\|\mathbf{v}\|^2}
\end{equation}
and
\begin{equation}
\sum_{\mathbf{v}\in\Lambda}e^{-\pi s^{-1}\|\mathbf{v}\|^2}\leq s^{\frac{n}{2}}\cdot\sum_{\mathbf{v}\in \Lambda}e^{-\pi\|\mathbf{v}\|^2},\ \text{for}\ s\geq1.
\end{equation}

From (\ref{bc1}), it is natural to see that the sampling probability for any specific $\mathbf{z}$ is lower bounded by the function $f(\sigma)$.
Furthermore, the derivative of function $f(\sigma)$ with respect to $\sigma\geq1/\sqrt{2\pi}$ is derived as follows
\begin{eqnarray}
\frac{\partial f(\sigma)}{\partial\sigma}=\frac{\left(n\sigma^{2}-\left\Vert \bar{\mathbf{B}}\mathbf{z}-\mathbf{c}\right\Vert ^{2}\right)\exp\left(-\frac{\left\Vert \bar{\mathbf{B}}\mathbf{z}-\mathbf{c}\right\Vert ^{2}}{2\sigma^{2}}\right)}{\sigma^{n+3}\left(\sqrt{2\pi}\right)^{n}}.
\label{bc2}
\end{eqnarray}

Subsequently, let the above derivative be zero, the optimized $\sigma$ that maximizes $f(\sigma)$ is obtained as
\begin{equation}
\sigma=\max\left\{\frac{\parallel \overline{\mathbf{B}}\mathbf{z}-\mathbf{c} \parallel}{\sqrt{n}}, \frac{1}{\sqrt{2\pi}}\right\},
\label{bc3}
\end{equation}
which implies that $\sigma$ should vary with $\parallel \overline{\mathbf{B}}\mathbf{z}-\mathbf{c} \parallel$ for a large lower bound of $\pi(\mathbf{z})$.

Clearly, the existence of the lower bound for $\pi(\mathbf{z})$ guarantees a reliable sampling probability of $\mathbf{z}$, which could be further optimized by the careful selection of $\sigma$.
Meanwhile, the requirement of $\sigma\geq1/\sqrt{2\pi}$ serves as a baseline to ensure the Markov chain evolves dynamically, even though the sampling probability below $\sigma=1/\sqrt{2\pi}$ seems rather attractive.

Hence, as for the target point $\mathbf{z}_{\text{cvp}}$ for lattice decoding, the choice of $\sigma$ due to (\ref{bc3}) turns out to be
\begin{equation}
\sigma_{\text{cvp}}=\max\left\{\frac{\parallel \overline{\mathbf{B}}\mathbf{z}_{\text{cvp}}-\mathbf{c} \parallel}{\sqrt{n}}, \frac{1}{\sqrt{2\pi}}\right\}.
\label{bc34}
\end{equation}
Generally speaking, regarding to different configurations of $\mathbf{B}$ and $\mathbf{w}$, such a flexible setting of $\sigma_{\text{cvp}}$ is more beneficial to the sampler decoding by providing a specific rather than statistic choice. For small value of $\|\overline{\mathbf{B}}\mathbf{z}_{\text{cvp}}-\mathbf{c}\|$, $\sigma_{\text{cvp}}$ tends to get smaller since $\mathbf{c}$ appears close to the lattice and vice versa, thus adaptively guiding the choice of $\sigma$ for each $\mathbf{z}_{\text{cvp}}$.

\begin{figure*}[t]
\centering\includegraphics[width=6in]{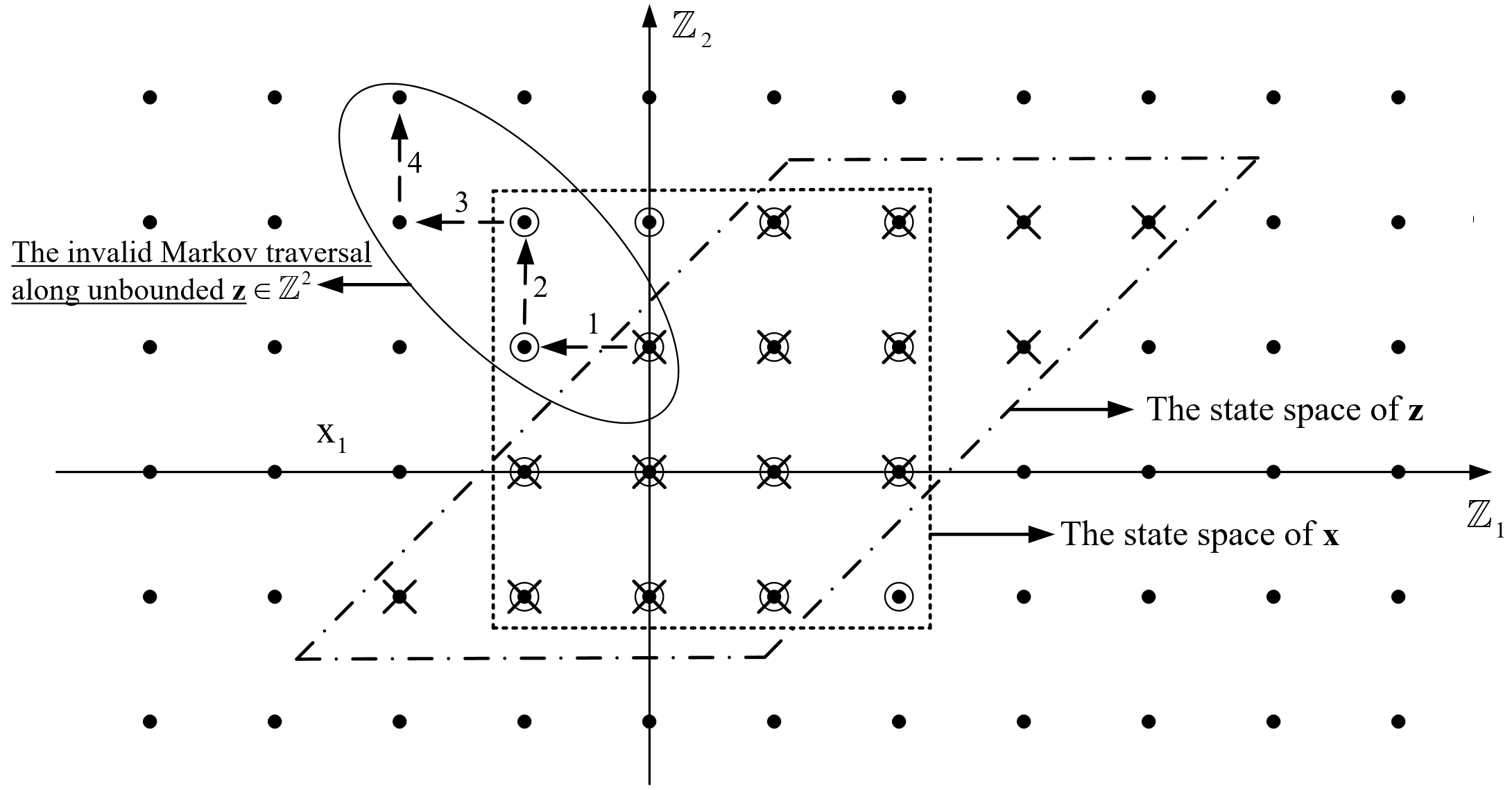}
\caption{Illustration of the defective Markov moves along with the unbounded state space $\mathbf{z}=\mathbf{U}^{-1}\mathbf{x}$. The original symbols $\mathbf{x}\in\mathcal{X}^2$ with $\mathcal{X}=\{-1,0,1,2\}$ and $U=[0\ 1; 1 -1]$, and the index $1,2,3,4$ along the arrows stand for the possible defective Markov moves.}
\label{IDD figure}
\end{figure*}

Unfortunately, it is impossible to get $\mathbf{z}_{\text{cvp}}$ for $\sigma_{\text{cvp}}$. Therefore, in practice, the initial starting point $\mathbf{z}^0$ can be applied as an approximation. Clearly, the closer of $\mathbf{z}^0$ to $\mathbf{z}_{\text{cvp}}$, the more accurate of the selected $\sigma$. This essentially poses a stringent request for the selection of $\mathbf{z}^0$. Fortunately, thanks to the lattice reduction, the required high quality initial starting point in lattice-reduction-aided Gibbs sampling can be guaranteed.
In this paper, the classic Babai's nearest plane algorithm (also known as successive interference cancelation (SIC) in MIMO detection) is utilized by
\begin{equation}
\mathbf{z}^0=\mathbf{z}_{\text{lll-sic}},
\end{equation}
where the decoding of $\mathbf{z}_{\text{lll-sic}}$ can be executed during the transformation from $D_{\mathcal{L}(\mathbf{B}),\sigma,\mathbf{c}}(\mathbf{x})$ to $D_{\mathcal{L}(\overline{\mathbf{B}}),\sigma,\mathbf{c}}(\mathbf{z})$.
To summarize, we reformat the proposed standard deviation as
\begin{equation}
\sigma_{\text{distance}}=\max\left\{\frac{\parallel \overline{\mathbf{B}}\mathbf{z}_{\text{lll-sic}}-\mathbf{c} \parallel}{\sqrt{n}}, \frac{1}{\sqrt{2\pi}}\right\}.
\label{bc3467}
\end{equation}
Again, we emphasize that other decoding schemes are also applicable to output $\mathbf{z}^0$ while the decoding performance improves with the accuracy of the approximation.

Besides the sampling probability, the initial starting point also plays an important role in the Markov mixing. More specifically, for the small set $C=\{\mathbf{x}: V(\mathbf{x})=\pi(\mathbf{x})^{-c}\leq d, c>0\}$ and $d>2b/(1-\lambda)$, the geometric ergodicity Markov chains will converge exponentially to the stationary distribution $\pi(\mathbf{x})$ as \cite{RosenthalMinorization}
\begin{equation}
\|\hspace{-.1em}P^n\hspace{-.1em}(\mathbf{x}^0, \hspace{-.2em}\cdot)-\pi(\cdot)\hspace{-.1em}\|_{TV}\hspace{-.2em}\leq\hspace{-.2em}(1\hspace{-.2em}-\hspace{-.1em}\delta)^{rn}\hspace{-.3em}+\hspace{-.1em}\left(\hspace{-.2em}\frac{U^r}{\alpha^{1-r}}\hspace{-.2em}\right)^{\hspace{-.2em}n}\hspace{-.4em}\left(\hspace{-.3em}1\hspace{-.2em}+\hspace{-.2em}\frac{b}{1\hspace{-.2em}-\hspace{-.2em}\lambda}\hspace{-.2em}+\hspace{-.2em}V\hspace{-.2em}(\mathbf{x}^0)\hspace{-.4em}\right)\hspace{-.3em},
\label{ddddda}
\end{equation}
where $0<r<1$, $0<\lambda<1$, $U=1+2(d+b)$ and $\alpha=\frac{1+d}{1+2b+\lambda d}$. From (\ref{ddddda}), starting the Markov chain with $\mathbf{z}^0$ as close to the center of the lattice Gaussian distribution (i.e., the query point $\mathbf{c}$) as possible would be a judicious choice for the efficient $t_{\text{mix}}$, which is accordance with our suggestion.

On the other hand, since $\mathbf{w}$ in (\ref{eqn:ML Decoding}) entails the additive white Gaussian noise (AWGN) with zero mean and variance $\sigma_w^2$, it follows that
\begin{equation}
\|\overline{\mathbf{B}}\mathbf{z}-\mathbf{c}\|^2=\|\mathbf{Bx}-\mathbf{c}\|^2\approx n \sigma_w^2
\label{bc6}
\end{equation}
by the \emph{law of large numbers}. Then, by simply substituting (\ref{bc6}) into (\ref{bc3}), the choice of $\sigma$ can be obtained in a statistic way, that is,
\begin{equation}
\sigma_{\text{statistic}}=\max\left\{\sigma_w, \frac{1}{\sqrt{2\pi}}\right\}.
\label{mnnm1}
\end{equation}
Interestingly, we point out that $\sigma=\sigma_w$ is just the conventional wisdom that is widely accepted by related works. However, compared to $\sigma_{\text{statistic}}$, it severely suffers from the \emph{stalling problem} as $\sigma^2_w$ shrinks intensively with the increase of SNR. Therefore, the lower bound $\sigma\geq1/\sqrt{2\pi}$ serves as a necessary complement to active the sampling away from the frozen status.
Note that the consistency behind choices of $\sigma_{\text{statistic}}$ and $\sigma_w$ suggests our analysis based on the sampling probability is tight enough, and we then advance it to more specific cases.

\subsection{Startup Mechanism based on Correct Decoding Radius $R$}
The application of the initial starting point arises a natural question: whether Gibbs sampling is necessary to every decoding case? In what follows, we try to answer this question from the perspective of correct decoding radius of BDD.

Theoretically, BDD targets at solving the decoding problem when the query point is close to the lattice within a certain distance, which corresponds to a restricted variant of CVP.
In BDD, the concept of correct decoding radius $R$ was proposed to serve as a benchmark for evaluating the decoding performance \cite{EmbeddingLuzzi}. More specifically, CVP is guaranteed to be solved if the distance between the query point $\mathbf{c}$ and the lattice $\Lambda$ (i.e., $d(\Lambda, \mathbf{c})$) is less than $R$.
As for Babai's nearest plane algorithm, its correct decoding radius is given by \cite{EmbeddingLuzzi}
\begin{equation}
R_{\text{lll-sic}}=\frac{1}{2}\min_{1\leq i\leq n}\|\widehat{\overline{\mathbf{b}}}_i\|.
\end{equation}
Here, we highlight the significance of LLL reduction again as it greatly increases $\min_i\|\widehat{\overline{\mathbf{b}}}_i\|$ compared to $\min_i\|\widehat{\mathbf{b}}_i\|$.
Furthermore, it has been shown in \cite{EmbeddingLuzzi} that $R_{\text{lll-sic}}$ is lower bounded as
\begin{equation}
R_{\text{lll-sic}}\geq\frac{1}{2\sqrt{n}\beta^{\frac{n-1}{4}}}\lambda_1(\mathbf{B}),
\label{eqn:ZF and SIC Decoding Radius}
\end{equation}
where $\beta=1/(\delta-1/4)$ and $\lambda_1$ denotes the \emph{minimum distance} of the lattice $\mathcal{L}(\mathbf{B})$.
Therefore, for the consideration of decoding efficiency, the correct decoding radius $R_{\text{lll-sic}}$ can be applied as a theoretical judgement to make the decision whether invoke Gibbs sampling or not. This means substantial decoding complexity will be saved without performance loss.

In particular, let $\mathbf{z}^0=\mathbf{z}_{\text{lll-sic}}$, the startup mechanism works based on the threshold $R_{\text{lll-sic}}$. If $\|\overline{\mathbf{B}}\mathbf{z}^0-\mathbf{c}\|\leq R_{\text{lll-sic}}$,
then there is no need to recall Gibbs sampler as $\mathbf{z}_{\text{cvp}}=\mathbf{z}_{\text{lll-sic}}$ for sure. Otherwise, Gibbs sampler is activated for a better decoding performance.
Moreover, such a judgement can be further relaxed with a constant $\alpha\geq1$
\begin{equation}
\|\overline{\mathbf{B}}\mathbf{z}^0-\mathbf{c}\|\leq\frac{\alpha}{2}\min_{1\leq i\leq n}\|\widehat{\overline{\mathbf{b}}}_i\|,
\label{v1a}
\end{equation}
which leads to a flexible trade-off between decoding performance and efficiency.

From the perspective of efficient sampler decoding, the need for the high quality initial starting point $\mathbf{z}^0$ is also in strong demand for providing a large size of correct decoding radius $R$. In essence, those demands actually reveal a salient feature of geometric ergodicity: the selection of the initial starting point is an indispensable part of the Markov mixing, which is worth to be well studied.
Here, we use it to work for the choice of $\sigma$, the pursuit of convergence as well as the startup mechanism, and we believe our work is just the tip of the iceberg. Meanwhile, the proposed startup mechanism based on the correct decoding radius also provides an adaptive strategy for other lattice decoding schemes especially in tackling with high-dimensional scenarios.


%
%

%



\begin{figure}[t]
\includegraphics[width=3.5in]{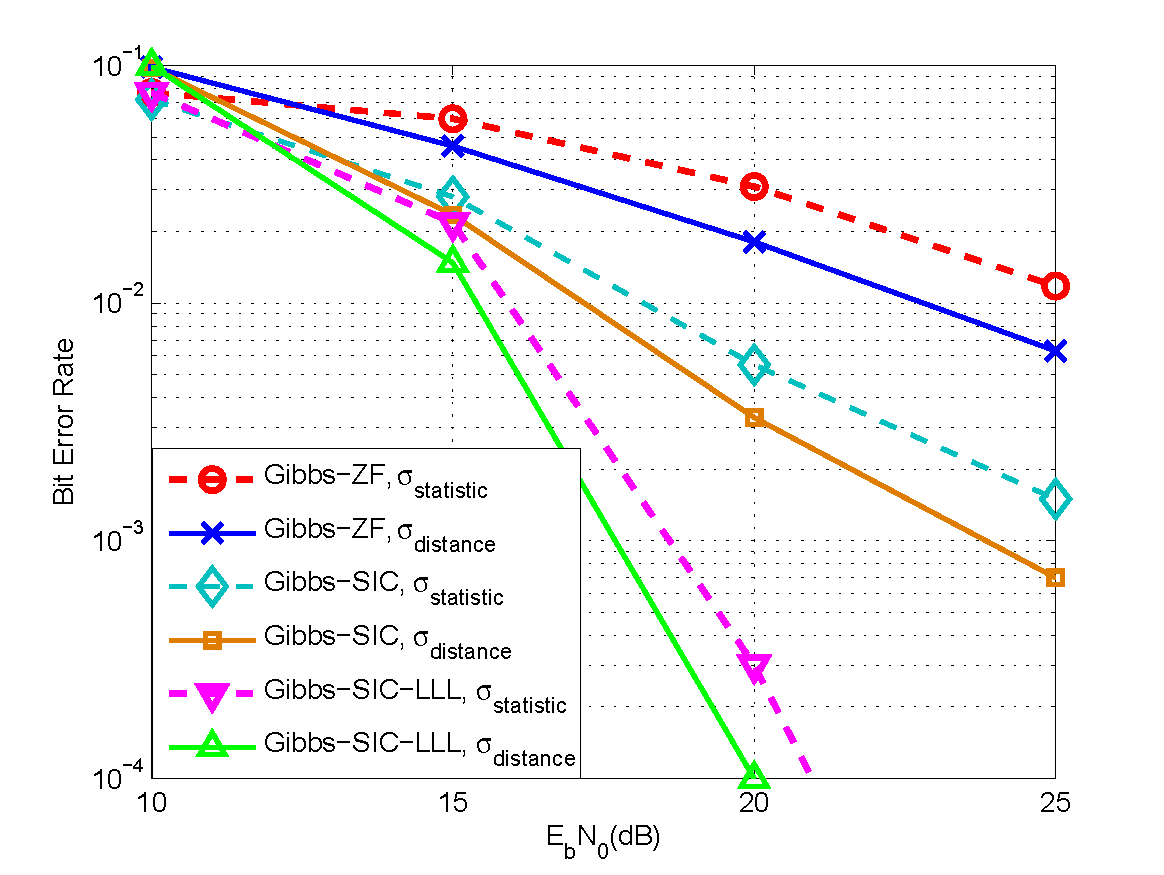}
\vspace{-1em}
  \caption{Bit error rate versus the number of iterations for the uncoded $12 \times 12$ MIMO system using 16-QAM.}
  \label{simulation 1}
\end{figure}

\subsection{Adoption for Large-Scale MIMO Detection}
In lattice decoding, both $\mathbf{x}$ and $\mathbf{z}$ have the same state space $\mathbb{Z}^n$. However, as for MIMO systems, the transmitted signal $\mathbf{x}\in \mathcal{X}^n$ normally belongs to a finite $M$-QAM constellation. Therefore, an inherent disadvantage associated with lattice reduction does exist since the state space of $\mathbf{z}=\mathbf{U}^{-1}\mathbf{x}$ after the nonlinear transformation is computationally expensive to get \cite{FiniteLattice}. Generally, in lattice-reduction-aided detection for MIMO systems, there are two suboptimal remedies to alleviate this problem. The first one directly discards those out-of-region points, which is referred to as naive lattice decoding (NLD)\cite{NLDTaherzadeh}. Another remedy is to restrict $\widehat{\mathbf{x}}=\mathbf{Uz}$ to the original set $\mathcal{X}^n$, which is commonly accepted \cite{WubbenLRMagzine}.

Unfortunately, both of these two remedies are not compatible with lattice-reduction-aided Gibbs sampling for MIMO detection as they only focus on the restriction of $\mathbf{x}$ at the final decision stage.
In sharp contrast to them, after the transformation by lattice reduction, the Markov mixing along $\mathbf{z}$ requires a clear state space, otherwise the Markov chain is going to be invalid due to the wild mixing.
Although approximation for the state space of $\mathbf{z}$ can be roughly made, it does not exactly correspond to the original state space of $\mathbf{x}$, leading to a defective Markov mixing. For a better understanding,
Fig. \ref{IDD figure} is presented as an illustration.

The essential reason behind such a problem is due to the acceptance mechanism of univariate sampling, i.e., every sampling candidate is accepted without any extra judgement or restriction. This actually raises a stringent requirement about the state space of Gibbs sampling since even a tiny disorder at the beginning would lead to a terrible error propagation along the mixing.
As a comparison, the MCMC based independent Metropolis-Hastings-Klein (MHK) algorithm utilizes an acceptance ratio to decide whether to admit the sample candidate or not\footnote{In principle, Gibbs sampling can be viewed as a special case of Metropolis-Hastings algorithm with acceptance ratio $p\equiv1$.}, and LLL reduction has been well adopted to it for a better decoding performance in MIMO detection \cite{ZhengWangTIT17}. To this end, in MIMO detection, lattice reduction without clear state space of $\mathbf{z}=\mathbf{U}^{-1}\mathbf{x}$ is not recommended to participate in the Markov mixing. This is in line with the observations from \cite{BaiLin1}, but it attributes the incompatibility to the increment of local minima by lattice reduction, which actually also exists in the lattice-reduction-aided detection.

Nevertheless, lattice reduction still works for Gibbs sampling in MIMO detection as a preprocessing stage to output the required initial starting point $\mathbf{x}^0=\mathbf{x}_{\text{lll-sic}}$.
Meanwhile, it is also easy to verify that our analysis about the choice of $\sigma$ as well as the startup mechanism for cases of $\mathbf{z}$ suit well for cases of $\mathbf{x}$ (simple scaling and shifting with respect to $\mathbf{x}$ are necessary to make it continuous integer). Therefore, given $\mathbf{x}^0=\mathbf{x}_{\text{lll-sic}}$, considerable performance gain and complexity reduction can be achieved.


\section{Simulation}
In this section, the performance of the proposed Gibbs sampling is evaluated in the large-scale MIMO detection.
Specifically, the $i$th entry of the transmitted signal $\mathbf{x}$, denoted as $x_i$, is a modulation symbol taken independently from an $M$-QAM constellation $\mathcal{X}$
with Gray mapping. Meanwhile, we assume a flat fading environment, where the square channel matrix
$\mathbf{H}$ contains uncorrelated complex Gaussian fading gains with unit
variance and remains constant over each frame duration. Let $E_b$ represents the average power per bit at the receiver, then the signal-to-noise ratio (SNR) $E_b/N_0=n/(\text{log}_2(M)\sigma_w^2)$ where $M$ is the modulation level and $\sigma_w^2$ is the noise variance. Then, we can express the system model as
\begin{equation}
\mathbf{c}=\mathbf{H}\mathbf{x}+\mathbf{w}.
\label{eqn:System Model3}
\end{equation}
Clearly, this decoding problem of $\widehat{\mathbf{x}}=\underset{\mathbf{x}\in \mathcal{X}^{n}}{\operatorname{arg~min}} \, \|\mathbf{H}\mathbf{x}-\mathbf{c}\|^2$ can be solved by sampling over the discrete Gaussian distribution
\begin{equation}
P_{\mathcal{L}(\mathbf{H}),\sigma,\mathbf{c}}(\mathbf{x})=\frac{e^{-\frac{1}{2\sigma^2}\parallel \mathbf{Hx}-\mathbf{c} \parallel^2}}{\sum_{\mathbf{x} \in \mathcal{X}^n}e^{-\frac{1}{2\sigma^2}\parallel \mathbf{Hx}-\mathbf{c} \parallel^2}}.
\label{discrete gaussian distribution}
\end{equation}

\begin{figure}[t]
\includegraphics[width=3.5in]{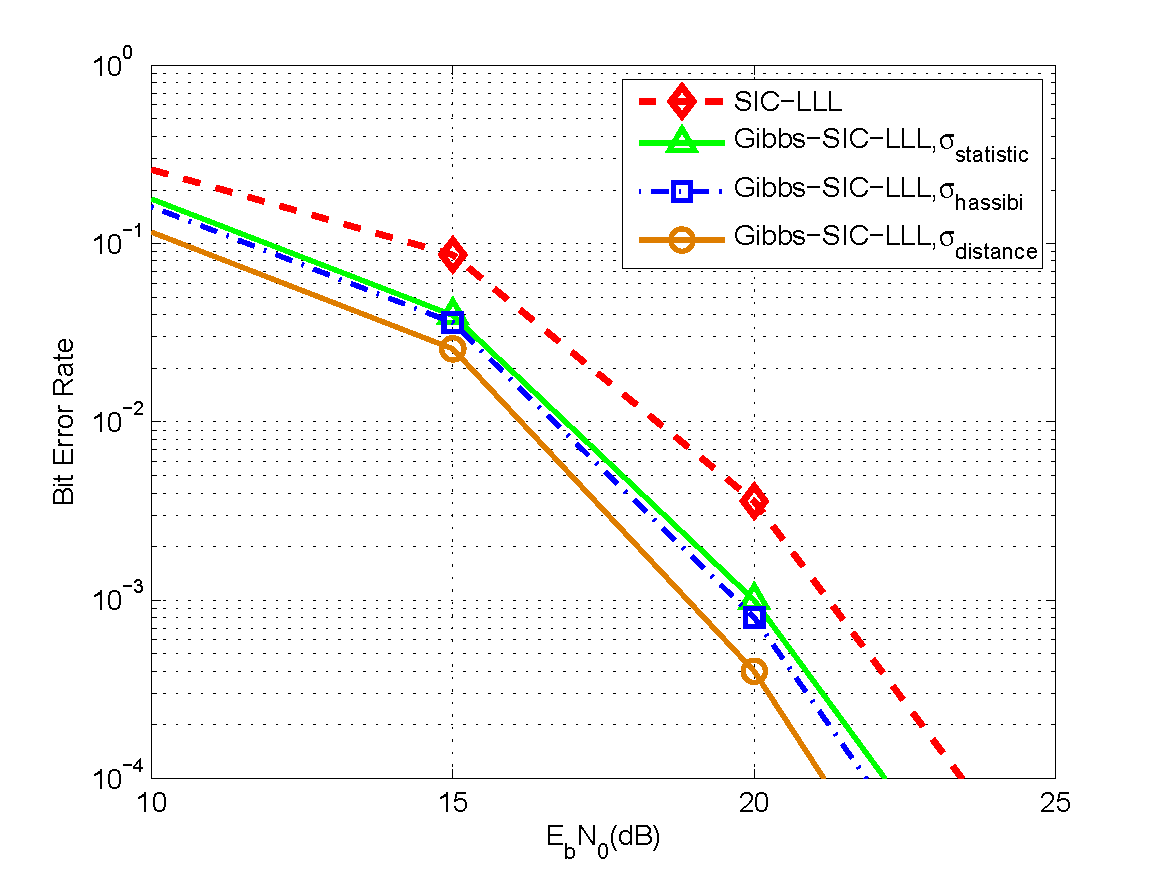}
\vspace{-1em}
  \caption{Bit error rate versus the number of iterations for the uncoded $16 \times 16$ MIMO system using 16-QAM.}
  \label{simulation 22}
\end{figure}

Fig. \ref{simulation 1} shows the bit error rate (BER) of Gibbs sampling detectors in a $12\times12$ uncoded MIMO system with 16-QAM, where all the samples generated by Gibbs sampling are taken into account for decoding. This corresponds to a lattice decoding scenario with restricted state space in dimension $n=24$. The systematic scan Gibbs algorithm performs 1-dimensional conditional sampling of $x_i$ in a backward order, thus completing a Markov move by a full iteration. As a comparison, Gibbs sampling with different initial starting points (i.e., outputted by ZF, SIC and SIC-LLL detectors respectively) are illustrated under the same Markov moves (i.e., $t=50$). Clearly, the decoding performance of Gibbs sampling based on SIC-LLL is the best, followed by the Gibbs sampling based on SIC and ZF. In terms of the convergence behaviour of geometric ergodicity, this is because the initial starting point provided by SIC-LLL is closest to the center of the distribution shown in (\ref{discrete gaussian distribution}), thus enabling a most efficient Markov mixing. Note that because of the distortion invoked by the nonlinear transformation of lattice reduction, in small $E_b/N_0$ regions, the decoding output by SIC-LLL may be even worse than that of standard SIC, which means SIC is more preferable in those cases.

\begin{figure}[t]
\includegraphics[width=3.5in]{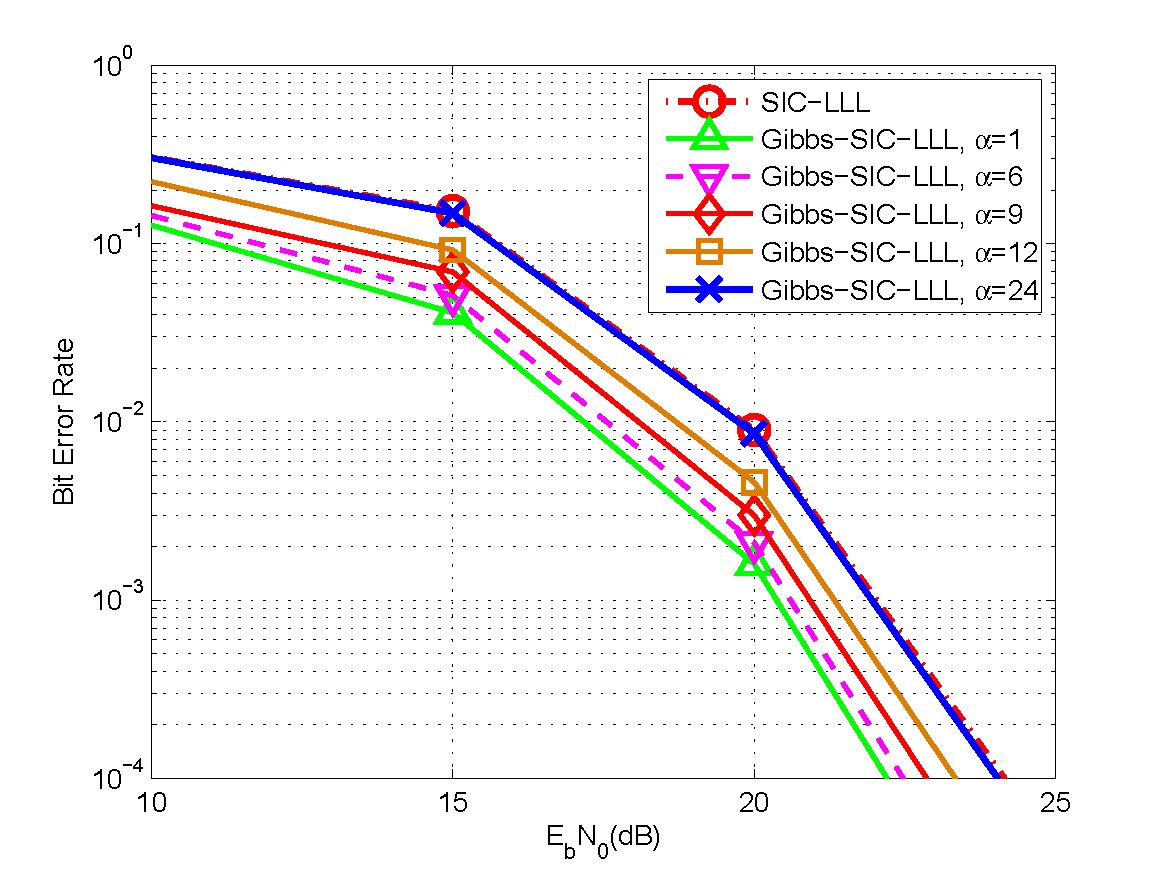}
\vspace{-1em}
  \caption{Bit error rate versus the number of iterations for the uncoded $24 \times 24$ MIMO system using 16-QAM.}
  \label{simulation 2}
\end{figure}

On the other hand, based on the approximation offered by the initial starting point, all the Gibbs sampling schemes with choices of $\sigma_{\text{distance}}$ given in (\ref{bc3467})
outperform their counterparts with $\sigma_{\text{statistic}}$ (the choice $\sigma=\sigma_w$ is contained in $\sigma_{\text{statistic}}$) shown in (\ref{mnnm1}). This clearly confirms our analysis as the former fully takes advantage of each specific decoding while the latter only offers a general solution by statistics.
In addition, for the sake of stalling problem, we point out the importance of the lower bound $\sigma\geq1/\sqrt{2\pi}$ contained in both $\sigma_{\text{distance}}$ and $\sigma_{\text{statistic}}$, which prevents the Markov mixing from getting frozen. For more details, the choice of $\sigma$ from \cite{HassibiMCMCnew} is added to Fig. \ref{simulation 22} in a $16\times16$ MIMO systems using 16-QAM.

In particular, Hassibi suggested
\begin{equation}
\sigma^2_{\text{hassibi}}=\frac{\text{SNR}}{\ln n}+\sqrt{\left(\frac{\text{SNR}}{\ln n}\right)^2-2\frac{\text{SNR}}{\ln n}},
\label{sigma_hassibi}
\end{equation}
which still belongs to a general solution of $\sigma$. From Fig. \ref{simulation 22}, compared to the basic SIC-LLL detection, considerable performance gain is obtained by Gibbs sampling detectors with $t=50$.
As expected, the choice of $\sigma_{\text{distance}}$ is the best due to its advantages of customized strategy, and the solution $\sigma_{\text{hassibi}}$ is slightly better than $\sigma_{\text{statistic}}$.
Additionally, under the help of LLL reduction, all the Gibbs sampling detectors achieve the full receive diversity gain.

\begin{figure}[t]
\includegraphics[width=3.5in]{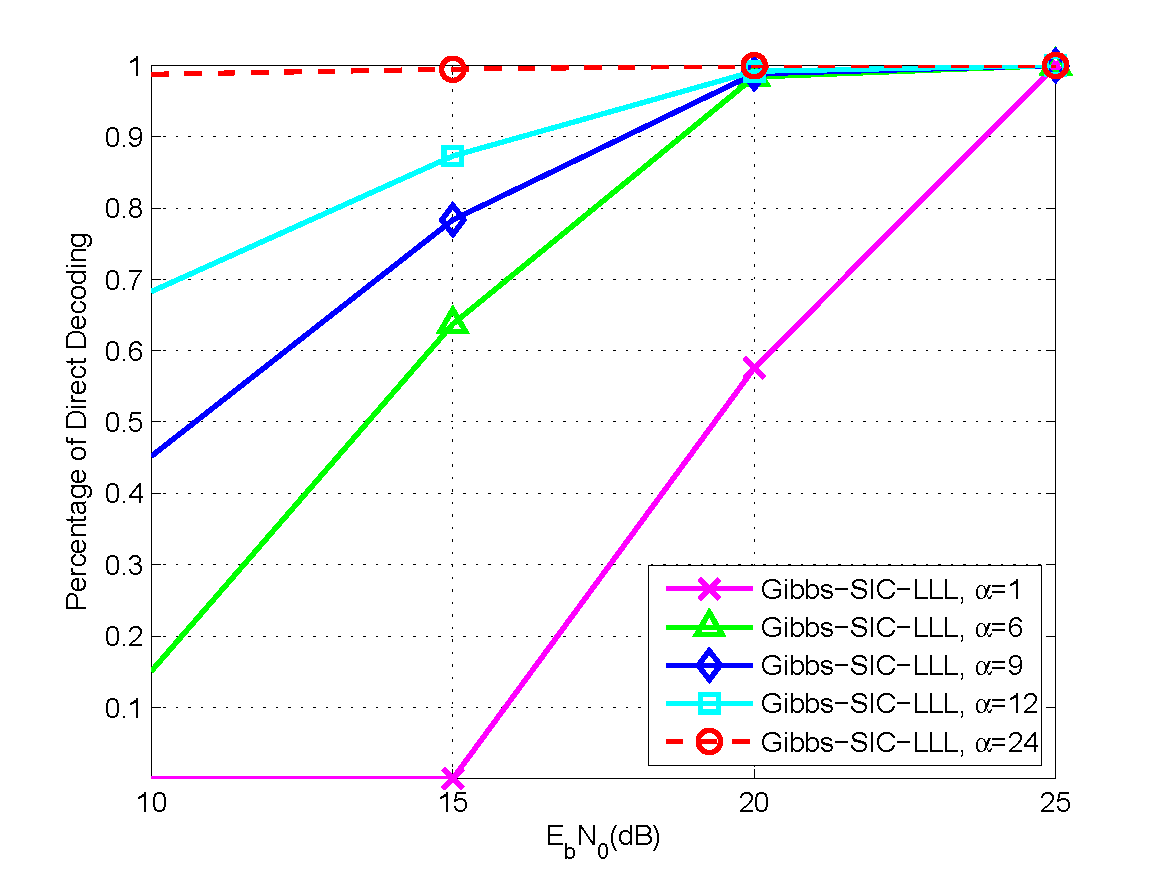}
\vspace{-1em}
  \caption{Bit error rate versus the percentage of direct decoding due to the startup mechanism for the uncoded $24 \times 24$ MIMO system using 16-QAM.}
  \label{simulation 3}
\end{figure}

In Fig. \ref{simulation 2}, the BERs of Gibbs sampling detectors with different coefficients $\alpha\geq1$ by means of correct decoding radius $R_{\text{lll-sic}}$ are evaluated in a $24\times24$ uncoded MIMO system with 16-QAM. The choice of $\sigma_{\text{distance}}$ is applied and the number of Markov moves is set by $t=50$. Specifically, given the initial starting point $\mathbf{x}_{0}=\mathbf{x}_{\text{lll-sic}}$, $\|\mathbf{H}\mathbf{x}_{\text{sic-lll}}-\mathbf{c}\|\leq\alpha R_{\text{lll-sic}}$ serves as a judgement to decide whether to invoke Gibbs sampling detector or not. As shown in Fig. \ref{simulation 2}, the decoding performance degrades gradually with the increase of $\alpha$, where $\alpha=1$ strictly obeys $\mathbf{x}_{\text{cvp}}=\mathbf{x}_{\text{lll-sic}}$ and $\alpha>1$ is a loose version of it. Clearly, with a moderate $\alpha$ (experimentally $n/8$), the decoding shows negligible performance loss, but saving considerable computational complexity from it. To be more precisely, the percentage of the direct decoding finished by the initial starting point, i.e., $\mathbf{x}_{\text{output}}=\mathbf{x}_{\text{sic-lll}}$, is depicted in Fig. \ref{simulation 3}.

Here, we highlight two salient features observed from Fig. \ref{simulation 3}. On one hand, with the increase of $E_b/N_0$, the noises are suppressed gradually, which significantly improves the quality of the output from SIC-LLL. In other words, more and more initial starting points are eligible to be directly outputted along with diminishing noises. Hence, from the perspective of each specific decoding point, the demand for Gibbs sampling detector should decrease along with SNR, which emphasizes the significance of the proposed startup mechanism by removing amounts of unnecessary sampling operations.
On the other hand, with the increase of $\alpha$, the startup judgement becomes loose while more initial starting points are allowed to output. This naturally leads to inevitable performance degradation. However, as shown in Fig. \ref{simulation 2}, a moderate choice of $\alpha$ (e.g. $\alpha=n/8$) still could achieve a promising trade-off between performance loss and complexity reduction.

\section{Conclusion}
In this paper, lattice reduction was introduced to Gibbs sampler for lattice Gaussian sampling.
The convergence rate of systematic scan Gibbs sampling was investigated in full details.
As demonstrated, the HGR maximal correlation over elements in lattice Gaussian sampling plays an indispensable role in the Markov mixing.
Therefore, lattice reduction is applied to establish an equivalent lattice Gaussian distribution, where Gibbs sampling can be carried out with a better convergence performance.
After that, we show that the proposed lattice-reduction-aided Gibbs sampling can be easily used as a sampler decoding scheme to solve the CVP.
To balance the trade-off between Markov mixing and sampler decoding, the choice of the standard deviation during the sampling was studied, and a suboptimal $\sigma$ based on the initial starting point was given.
Finally, to pursuit an efficient sampler decoding, a startup mechanism resorting to correct decoding radius from BDD was proposed.

\section*{Acknowledgment}
The authors would like to thank Dr. Cong Ling (Imperial College London, UK) for his helpful discussions and insightful suggestions.

%
\bibliographystyle{IEEEtran}
\bibliography{IEEEabrv,reference1}

\begin{thebibliography}{10}
\providecommand{\url}[1]{#1}
\csname url@samestyle\endcsname
\providecommand{\newblock}{\relax}
\providecommand{\bibinfo}[2]{#2}
\providecommand{\BIBentrySTDinterwordspacing}{\spaceskip=0pt\relax}
\providecommand{\BIBentryALTinterwordstretchfactor}{4}
\providecommand{\BIBentryALTinterwordspacing}{\spaceskip=\fontdimen2\font plus
\BIBentryALTinterwordstretchfactor\fontdimen3\font minus
  \fontdimen4\font\relax}
\providecommand{\BIBforeignlanguage}[2]{{%
\expandafter\ifx\csname l@#1\endcsname\relax
\typeout{** WARNING: IEEEtran.bst: No hyphenation pattern has been}%
\typeout{** loaded for the language `#1'. Using the pattern for}%
\typeout{** the default language instead.}%
\else
\language=\csname l@#1\endcsname
\fi
#2}}
\providecommand{\BIBdecl}{\relax}
\BIBdecl

\bibitem{Banaszczyk}
W.~Banaszczyk, ``New bounds in some transference theorems in the geometry of
  numbers,'' \emph{Math. Ann.}, vol. 296, pp. 625--635, 1993.

\bibitem{Forney_89}
G.~Forney and L.-F. Wei, ``Multidimensional constellations--{Part II}: Voronoi
  constellations,'' \emph{IEEE J. Sel. Areas Commun.}, vol.~7, no.~6, pp.
  941--958, Aug. 1989.

\bibitem{Kschischang_Pasupathy}
F.~R. Kschischang and S.~Pasupathy, ``Optimal nonuniform signaling for
  {G}aussian channels,'' \emph{{IEEE} Trans. Inform. Theory}, vol.~39, pp.
  913--929, May. 1993.

\bibitem{LiuLing2}
L.~Liu and C.~Ling, ``Polar codes and polar lattices for independent fading
  channels,'' \emph{IEEE Transactions on Communications}, vol.~64, no.~12, pp.
  4923--4935, Dec 2016.

\bibitem{LB_13}
C.~Ling and J.-C. Belfiore, ``Achieiving the {AWGN} channel capacity with
  lattice {Gaussian} coding,'' \emph{IEEE Trans. Inform. Theory}, vol.~60,
  no.~10, pp. 5918--5929, Oct. 2014.

\bibitem{LiuLing1}
L.~Liu, Y.~Yan, and C.~Ling, ``Achieving secrecy capacity of the {G}aussian
  wiretap channel with polar lattices,'' \emph{IEEE Transactions on Information
  Theory}, vol.~64, no.~3, pp. 1647--1665, March 2018.

\bibitem{LLBS_12}
C.~Ling, L.~Luzzi, J.-C. Belfiore, and D.~Stehl\'{e}, ``Semantically secure
  lattice codes for the {G}aussian wiretap channel,'' \emph{IEEE Trans. Inform.
  Theory}, vol.~60, no.~10, pp. 6399--6416, Oct. 2014.

\bibitem{7360779}
H.~Mirghasemi and J.~C. Belfiore, ``Lattice code design criterion for mimo
  wiretap channels,'' in \emph{Proc IEEE Information Theory Workshop (ITW)},
  Oct 2015, pp. 277--281.

\bibitem{7058433}
S.~Vatedka, N.~Kashyap, and A.~Thangaraj, ``Secure compute-and-forward in a
  bidirectional relay,'' \emph{IEEE Transactions on Information Theory},
  vol.~61, no.~5, pp. 2531--2556, May 2015.

\bibitem{MicciancioGaussian}
D.~Micciancio and O.~Regev, ``Worst-case to average-case reductions based on
  {Gaussian} measures,'' in \emph{Proc. Ann. Symp. Found. Computer Science},
  Rome, Italy, Oct. 2004, pp. 372--381.

\bibitem{Regevlearning}
O.~Regev, ``On lattice, learning with errors, random linear codes, and
  cryptography,'' \emph{J. ACM}, vol.~56, no.~6, pp. 34:1--34:40, 2009.

\bibitem{GentryDissertation}
C.~Gentry, ``A fully homomorphic encryption scheme,'' Ph.D. dissertation,
  Stanford University, USA, 2009.

\bibitem{Trapdoor}
C.~Gentry, C.~Peikert, and V.~Vaikuntanathan, ``Trapdoors for hard lattices and
  new cryptographic constructions,'' in \emph{Proc. 40th Ann. ACM Symp. Theory
  of Comput.}, Victoria, Canada, 2008, pp. 197--206.

\bibitem{RegevSolvingtheShortestVectorProblem}
D.~Aggarwal, D.~Dadush, O.~Regev, and N.~Stephens-Davidowitz, ``Solving the
  shortest vector problem in $2^n$ time via discrete {Gaussian} sampling,''
  \emph{STOC}, 2015.

\bibitem{RegevSolvingtheClosestVectorProblem}
D.~Aggarwal, D.~Dadush, and N.~Stephens-Davidowitz, ``Solving the closest
  vector problem in $2^n$ time --- the discrete {Gaussian} strike again!''
  \emph{FOCS}, 2015.

\bibitem{DamenDetectionSearch}
M.~O. Damen, H.~E. Gamal, and G.~Caire, ``On maximum-likelihood detection and
  the search for the closest lattice point,'' \emph{IEEE Trans. Inform.
  Theory}, vol.~49, pp. 2389--2401, Oct. 2003.

\bibitem{CB16}
A.~Campello and J.-C. Belfiore, ``Sampling algorithms for lattice {G}aussian
  codes,'' in \emph{Proc. International Zurich Seminar on Communications
  (IZS)}, Zurich, Switzerland, 2016.

\bibitem{Klein}
P.~Klein, ``Finding the closest lattice vector when it is unusually close,'' in
  \emph{ACM-SIAM Symp. Discr. Algorithms}, 2000, pp. 937--941.

\bibitem{ZhengWangTIT15}
Z.~Wang and C.~Ling, ``On the geometric ergodicity of {M}etropolis-{H}astings
  algorithms for lattice {G}aussian sampling,'' \emph{IEEE Transactions on
  Information Theory}, vol.~64, no.~2, pp. 738--751, Feb. 2018.

\bibitem{ZhengWangTIT17}
------, ``Lattice {G}aussian sampling by {M}arkov chain {M}onte {C}arlo:
  Bounded distance decoding and trapdoor sampling,'' \emph{Submitted to IEEE
  Transactions on Information Theory}, [Online]
  Available:http://arxiv.org/abs/1704.02673.

\bibitem{mixingtimemarkovchain}
D.~A. Levin, Y.~Peres, and E.~L. Wilmer, \emph{Markov Chains and Mixing Time},
  American Mathematical Society, 2008.

\bibitem{ZhengWangMCMCLatticeGaussian}
Z.~Wang, C.~Ling, and G.~Hanrot, ``{Markov chain Monte Carlo} algorithms for
  lattice {Gaussian} sampling,'' in \emph{Proc. IEEE International Symposium on
  Information Theory (ISIT)}, Honolulu, USA, Jun. 2014, pp. 1489--1493.

\bibitem{ZhengWangSysmetricGibbs}
Z.~Wang and C.~Ling, ``Symmetric {M}etropolis-within-{G}ibbs algorithm for
  lattice {G}aussian sampling,'' in \emph{Proc. IEEE Information Theory
  Workshop (ITW)}, Sept. 2016.

\bibitem{ITW2017}
------, ``On the geometric ergodicity of {G}ibbs algorithm for lattice
  {G}aussian sampling,'' in \emph{Proc. IEEE Information Theory Workshop
  (ITW)}, 2017, pp. 269--273.

\bibitem{TabuSrinidhi}
N.~Srinidhi, T.~Datta, A.~Chockalingam, and B.~S. Rajan, ``Layered {T}abu
  search algorithm for large-{MIMO} detection and a lower bound on {ML}
  performance,'' \emph{IEEE Transactions on Communications}, vol.~59, no.~11,
  pp. 2955--2963, Nov. 2011.

\bibitem{DaiL1}
L.~Dai, X.~Gao, X.~Su, S.~Han, C.~I, and Z.~Wang, ``Low-complexity soft-output
  signal detection based on {G}auss-{S}eidel method for uplink multiuser
  large-scale {MIMO} systems,'' \emph{IEEE Transactions on Vehicular
  Technology}, vol.~64, no.~10, pp. 4839--4845, Oct. 2015.

\bibitem{GaoX1}
A.~Lu, X.~Gao, Y.~R. Zheng, and C.~Xiao, ``Low complexity polynomial expansion
  detector with deterministic equivalents of the moments of channel {G}ram
  matrix for massive {MIMO} uplink,'' \emph{IEEE Transactions on
  Communications}, vol.~64, no.~2, pp. 586--600, Feb. 2016.

\bibitem{MSMIMO1}
S.~Wu, L.~Kuang, Z.~Ni, J.~Lu, D.~Huang, and Q.~Guo, ``Low-complexity iterative
  detection for large-scale multiuser {MIMO-OFDM} systems using approximate
  message passing,'' \emph{IEEE Journal of Selected Topics in Signal
  Processing}, vol.~8, no.~5, pp. 902--915, Oct. 2014.

\bibitem{ZZY1}
Z.~Zhang, X.~Cai, C.~Li, C.~Zhong, and H.~Dai, ``One-bit quantized massive
  {MIMO} detection based on variational approximate message passing,''
  \emph{IEEE Transactions on Signal Processing}, vol.~66, no.~9, pp.
  2358--2373, May. 2018.

\bibitem{HassibiMCMCnew}
B.~Hassibi, M.~Hansen, A.~Dimakis, H.~Alshamary, and W.~Xu, ``Optimized {Markov
  Chain Monte Carlo} for signal detection in {MIMO} systems: {An} analysis of
  the stationary distribution and mixing time,'' \emph{IEEE Transactions on
  Signal Processing}, vol.~62, no.~17, pp. 4436--4450, Sep. 2014.

\bibitem{McmcDatta}
T.~Datta, N.~Kumar, A.~Chockalingam, and B.~Rajan, ``A novel {Monte Carlo}
  sampling based receiver for large-scale uplink multiuser {MIMO} systems,''
  \emph{IEEE Transactions on Vehicular Technology,}, vol.~62, no.~7, pp.
  3019--3038, Sep. 2013.

\bibitem{XiaodongWangMultilevel}
P.~Aggarwal and X.~Wang, ``Multilevel sequential {M}onte {C}arlo algorithms for
  {MIMO} demodulation,'' \emph{IEEE Transactions on Wireless Communications},
  vol.~6, no.~2, pp. 750--758, Feb. 2007.

\bibitem{MCMCHaidongZhu}
H.~Zhu, B.~Farhang-Boroujeny, and R.-R. Chen, ``On performance of sphere
  decoding and {Markov} chain {Monte} {Carlo} detection methods,'' \emph{IEEE
  Signal Processing Letters}, vol.~12, no.~10, pp. 669--672, 2005.

\bibitem{ChoiMCMC1}
J.~Choi, ``An {MCMC-MIMO} detector as a stochastic linear system solver using
  successive overrelexation,'' \emph{IEEE Transactions on Wireless
  Communications}, vol.~15, no.~2, pp. 1445--1455, Feb. 2016.

\bibitem{MCMCBehrouz}
B.~Farhang-Boroujeny, H.~Zhu, and Z.~Shi, ``Markov chain {Monte Carlo}
  algorithms for {CDMA} and {MIMO} communication systems,'' \emph{IEEE Trans.
  Signal Process.}, vol.~54, no.~5, pp. 1896--1909, 2006.

\bibitem{BaiLin1}
L.~Bai, T.~Li, J.~Liu, Q.~Yu, and J.~Choi, ``Large-scale {MIMO} detection using
  {MCMC} approach with blockwise sampling,'' \emph{IEEE Transactions on
  Communications}, vol.~64, no.~29, pp. 3697--3707, Sept. 2006.

\bibitem{ChenMCMC}
R.~Chen, J.~Liu, and X.~Wang, ``Convergence analysis and comparisons of
  {Markov} {chain} {Monte Carlo} algorithms in digital communications,''
  \emph{IEEE Trans. on Signal Process.}, vol.~50, no.~2, pp. 255--270, 2002.

\bibitem{ChoiMCMC2}
J.~C. Hedstrom, C.~H. Yuen, R.~Chen, and B.~Farhang-Boroujeny, ``Achieving near
  {MAP} performance with an excited {M}arkov chain {M}onte {C}arlo {MIMO}
  detector,'' \emph{IEEE Transactions on Wireless Communications}, vol.~16,
  no.~12, pp. 7718--7732, Dec. 2017.

\bibitem{WubbenMMSE}
D.~Wubben, R.~Bohnke, V.~Kuhn, and K.~D. Kammeyer, ``Near-maximum-likelihood
  detection of {MIMO} systems using {MMSE}-based lattice reduction,'' in
  \emph{Proc. IEEE Int. Conf. Commun.(ICC'04)}, Paris, France, Jun. 2004, pp.
  798--802.

\bibitem{Luo1}
K.~Luo and A.~Manikas, ``Joint transmitter¨creceiver optimization in
  multitarget {MIMO} radar,'' \emph{IEEE Transactions on Signal Processing},
  vol.~65, no.~23, pp. 6292--6302, Dec 2017.

\bibitem{xia2}
H.~Cheng, Y.~Xia, Y.~Huang, L.~Yang, and D.~P. Mandic, ``A normalized complex
  {LMS} based blind {I}/{Q} imbalance compensator for {GFDM} receivers and its
  full second-order performance analysis,'' \emph{IEEE Trans. on Signal
  Process.}, vol.~66, no.~17, pp. 4701--4712, Sep. 2018.

\bibitem{Wuqihui1}
Q.~Wu, G.~Ding, J.~Wang, and Y.~Yao, ``Spatial-temporal opportunity detection
  for spectrum-heterogeneous cognitive radio networks: Two-dimensional
  sensing,'' \emph{IEEE Transactions on Wireless Communications}, vol.~12,
  no.~2, pp. 516--526, Feb. 2013.

\bibitem{Zhuang1}
J.~Zhuang, H.~Xiong, W.~Wang, and Z.~Chen, ``Application of manifold separation
  to parametric localization for incoherently distributed sources,'' \emph{IEEE
  Transactions on Signal Processing}, vol.~66, no.~11, pp. 2849--2860, June
  2018.

\bibitem{Xiangmin1}
M.~Xiang, B.~S. Dees, and D.~P. Mandic, ``Multiple-model adaptive estimation
  for 3-{D} and 4-{D} signals: A widely linear quaternion approach,''
  \emph{IEEE Transactions on Neural Networks and Learning Systems}, pp. 1--13,
  2018.

\bibitem{CongRandom}
S.~Liu, C.~Ling, and D.~Stehl\'{e}, ``{Decoding by sampling: A randomized
  lattice algorithm for bounded distance decoding},'' \emph{IEEE Trans. Inform.
  Theory}, vol.~57, pp. 5933--5945, Sep. 2011.

\bibitem{Xia1}
Y.~Xia and D.~P. Mandic, ``Augmented performance bounds on strictly linear and
  widely linear estimators with complex data,'' \emph{IEEE Trans. on Signal
  Process.}, vol.~66, no.~2, pp. 507--514, Jan 2018.

\bibitem{DGStoCVPSVP}
N.~Stephens-Davidowitz, ``Discrete {Gaussian} sampling reduces to {CVP} and
  {SVP},'' submitted for publication. [Online]. Available:
  http://arxiv.org/abs/1506.07490.

\bibitem{DerandomizedJ}
Z.~Wang, S.~Liu, and C.~Ling, ``Decoding by sampling - {Part} {II}:
  Derandomization and soft-output decoding,'' \emph{IEEE Trans. Commun.},
  vol.~61, no.~11, pp. 4630--4639, Nov. 2013.

\bibitem{ScanorderHe}
B.~He, C.~D. Sa, I.~Mitliagkas, and C.~Re, ``Scan order in {G}ibbs sampling:
  {M}odels in which it matters and bounds on how much,'' \emph{Neural
  Information Processing Systems (NIPS)}, pp. 1--9, 2016.

\bibitem{RobertsGeneralstatespace}
G.~O. Roberts, ``General state space {Markov} chains and {MCMC} algorithms,''
  \emph{Probability Surveys}, vol.~1, pp. 20--71, 2004.

\bibitem{WubbenLRMagzine}
D.~Wubben, D.~Seethaler, J.~Jalden, and G.~Matz, ``Lattice reduction,''
  \emph{IEEE Signal Processing Magazine}, vol.~28, no.~3, pp. 70--91, May 2011.

\bibitem{Shanxiang1}
S.~Lyu and C.~Ling, ``Boosted {KZ} and {LLL} algorithms,'' \emph{IEEE
  Transactions on Signal Processing}, vol.~65, no.~18, pp. 4784--4796, Sept
  2017.

\bibitem{Adeane}
J.Adeane, M.~Rodrigues, and I.~Wassell, ``Lattice-reduction-aided detection for
  {MIMO-OFDM-CDM} communication systems,'' \emph{IET commun.}, vol.~1, pp.
  526--531, 2007.

\bibitem{LLLoriginal}
A.~K. Lenstra, H.~W. Lenstra, and L.~Lovasz, ``Factoring polynomials with
  rational coefficients,'' \emph{Math. Annalen}, vol. 261, pp. 515--534, 1982.

\bibitem{MTaherzadeh2007}
M.~Taherzadeh, A.~Mobasher, and A.~Khandani, ``{LLL} reduction achieves the
  receive diversity in {MIMO} decoding,'' \emph{IEEE Trans. Inform. Theory},
  vol.~53, pp. 4801--4805, Dec. 2007.

\bibitem{JaldenDMT}
J.~Jalden and P.~Elia, ``{LR}-aided {MMSE} lattice decoding is {DMT} optimal
  for all approximately universal codes,'' in \emph{Proc. IEEE Int. Symp.
  Inform. Theory}, Seoul, Korea, July. 2009, pp. 1263--1267.

\bibitem{CongEffective}
C.~Ling and N.~Howgrave-Graham, ``Effective {LLL} reduction for lattice
  decoding,'' in \emph{Proc IEEE Int. Symp. Inform. Theory}, Nice, France, Jun.
  2007, pp. 196--200.

\bibitem{MaximalCorrelation2}
H.~O. Hirschfeld, ``A connection between correlation and contingency,'' in
  \emph{Proc. Cambridge Philosophical}, 1935, pp. 520--524.

\bibitem{MaximalCorrelation3}
H.~Gebelein, ``Das statistische problem der korrelation als variations- und
  eigenwert-problem und sein zusammenhang mit der ausgleichungsrechnung,''
  \emph{Zeitschrift fur angew. Math. und Mech.}, vol.~21, pp. 364--379, 1941.

\bibitem{MaximalCorrelation4}
A.~R\'enyi, ``On measures of dependence,'' \emph{Acta Math. Hung.}, vol.~10,
  pp. 441--451, 1959.

\bibitem{MaximalCorrelation1}
N.~Papadatos and T.~Xifara, ``A simple method for obtaining the maximal
  correlation coefficient and related characterizations,'' \emph{J. Multiv.
  Anal.}, vol. 118, pp. 102--114, 2013.

\bibitem{RobertsAndSahu}
G.~O. Roberts and S.~K. Sahu, ``Updating schemes, correlation structure,
  blocking and parameterization for {Gibbs} sampler,'' \emph{J. Roy. Statist.
  Soc. Series B}, \textbf{59}(2): 291-317, 1997.

\bibitem{YosidaBook}
K.~Yosida, \emph{Functional Analysis}.\hskip 1em plus 0.5em minus 0.4em\relax
  New York: Springer-Verlag, 6th ed., 1980.

\bibitem{LiuCovarianceSchemes}
J.~S. Liu and W.~H. Wong, ``Covariance structure of the {Gibbs} sampler with
  applications to the comparisons of estimators and augmentation schemes,''
  \emph{Biometrika}, \textbf{81}(1): 27-40, 1995.

\bibitem{LiuBook}
J.~S. Liu, \emph{Monte Carlo Strategies in Scientific Computing}, New York:
  Springer-Verlag, 2001.

\bibitem{Fill1991}
J.~A. Fill, ``Eigenvalue bounds on convergence to stationary for nonreversible
  markov chains, with application to the exclusion process.'' in \emph{Proc.
  Annals of Applied Probability}, vol.~1, 1991, pp. 62--87.

\bibitem{LiuCovarianceScans}
J.~S. Liu, W.~H. Wong, and A.~Kong, ``Covariance structure and convergence rate
  of the {Gibbs} sampler with various scans,'' \emph{J. Roy. Statist. Soc.
  Series B}, \textbf{57}(1): 157-169, 1995.

\bibitem{SpectralGapL2}
I.~Kontoyannis and S.~P. Meyn, ``Geometric ergodicity and spectral gap of
  non-reversible real valued {M}arkov chains,'' in \emph{Proc. Probab. Theory
  and related Fields}, vol. 154, 2012, pp. 327--339.

\bibitem{ParallelTempering2}
C.~J. Geyer and E.~A. Thompson, ``Annealing {M}arkov chain {M}onte {C}arlo with
  applications to ancestral inference,'' \emph{J. Amer. Statist. Assoc.},
  vol.~90, pp. 909--920, 1995.

\bibitem{RosenthalMinorization}
J.~S. Rosenthal, ``Minorization conditions and convergence rates for {Markov}
  chain {Monte} {Carlo},'' \emph{J. Amer. Statist. Assoc.}, vol.~90, pp.
  558--566, 1995.

\bibitem{EmbeddingLuzzi}
L.~Luzzi, D.~Stehl\'e, and C.~Ling, ``Decoding by embedding: correct decoding
  radius and {DMT} optimality,'' \emph{IEEE Trans. Inform. Theory}, vol.~59,
  no.~5, pp. 2960--2973, 2013.

\bibitem{FiniteLattice}
C.~Studer, D.~Seethaler, and H.~Bolcskei, ``Finite lattice-size effects in
  {MIMO} detection,'' in \emph{Proc. Asilomar Conference on Signals, Systems
  and Computers}, vol. 154, Oct. 2008, pp. 2032--2037.

\bibitem{NLDTaherzadeh}
M.~Taherzadeh and A.~K. Khandani, ``On the limitations of the naive lattice
  decoding,'' \emph{IEEE Transactions on Information Theory}, vol.~56, no.~10,
  pp. 4820--4826, 2010.

\end{thebibliography}

\end{document}